\documentclass[runningheads]{llncs}
\usepackage[utf8]{inputenc}
\usepackage{stmaryrd}
\usepackage{xcolor}
\usepackage{amsmath}
\usepackage{amsfonts}
\usepackage{bussproofs}
\usepackage{xspace}
\usepackage{listings}
\usepackage{graphicx}
\usepackage{placeins}
\usepackage[inline]{enumitem}

\newcommand{\driple}[2]{\genfrac{\{}{\}}{0pt}{}{#1}{#2}}
\newcommand{\sdriple}[2]{\{#1;~#2\}}
\newcommand{\rulename}[1]{\textbf{\scriptsize(\textsf{#1})}}
\newcommand{\ek}[1]{\marginpar{\textcolor{brown}{EK: #1}}}
\newcommand{\dg}[1]{\marginpar{\textcolor{blue}{DG: #1}}}
\newcommand{\EK}[1]{\textcolor{brown}{EK: #1}}
\newcommand{\DG}[1]{\textcolor{blue}{DG: #1}}

\newcommand{\state}{\sigma}
\newcommand{\states}{\mathtt{S}}

\newcommand{\statement}{s\xspace}
\newcommand{\liftspec}{\hat{\mu}}
\newcommand{\liftstate}{\mu}
\newcommand{\liftdirect}{\liftspec_\mathsf{direct}}
\newcommand{\abduct}{\alpha}
\newcommand{\statephi}{\Phi}
\newcommand{\liftphi}{\Delta}
\newcommand{\knowledge}{\mathbf{K}}
\newcommand{\contracts}{\mathbf{C}}
\newcommand{\many}[1]{\overline{#1}}

\newcommand{\setdef}[2]{\left\{ #1 \>\middle|\> #2 \right\}}
\newcommand{\denot}[1]{\left[\!\left[#1\right]\!\right]}

\newcommand{\denotCK}[1]{\left[\!\left[#1\right]\!\right]_{\contracts, \knowledge}}
\newcommand{\arexdenot}[1]{{\mathcal{A}\!\denot{#1}}}
\newcommand{\bexdenot}[1]{{\mathcal{B}\!\denot{#1}}}

\DeclareOldFontCommand{\rm}{\normalfont\rmfamily}{\mathrm}
\DeclareOldFontCommand{\sf}{\normalfont\sffamily}{\mathsf}
\DeclareOldFontCommand{\tt}{\normalfont\ttfamily}{\mathtt}
\DeclareOldFontCommand{\bf}{\normalfont\bfseries}{\mathbf}
\DeclareOldFontCommand{\it}{\normalfont\itshape}{\mathit}
\DeclareOldFontCommand{\sl}{\normalfont\slshape}{\@nomath\sl}
\DeclareOldFontCommand{\sc}{\normalfont\scshape}{\@nomath\sc}
\DeclareRobustCommand*\cal{\@fontswitch\relax\mathcal}
\DeclareRobustCommand*\mit{\@fontswitch\relax\mathnormal}

\colorlet{keywordcolor}{blue!50!black}
\colorlet{commentcolor}{green!60!black}
\colorlet{typecolor}{violet}
 
\newcommand{\sourcefont}{\ttfamily\small}
\newcommand{\commentfont}{\slshape\rmfamily\color{commentcolor}}

\lstdefinelanguage{Lang}{
        keywords={if,then,else,fi,while,do,od,record,var,proc,null,begin,end},
        keywordstyle=\bfseries\sffamily,
        sensitive=true,
        comment=[l]{//},
        morecomment=[s]{/*}{*/},
        morestring=[b]"
}

\lstdefinestyle{code}{
        basicstyle=\sourcefont\upshape,
        keywordstyle=\color{keywordcolor}\bfseries\sffamily,
        commentstyle=\commentfont,
        columns=fullflexible,
        mathescape=true,
        escapechar={\#},
        keepspaces=true,
        showstringspaces=false,
        aboveskip=8pt, 
        numbers=left,
        stepnumber=1, 
        numberstyle=\ttfamily\scriptsize\color{gray},
        numbersep=4pt,
        xleftmargin=1.5em,
        xrightmargin=1.5em,
        framexleftmargin=1.2em,
        framexrightmargin=1em,
        framextopmargin=0.5ex,
        breaklines=true,
        breakindent=3pt,
}

\lstdefinestyle{lang}{
        style=code,
        language=Lang,
}

\lstnewenvironment{langcode}[1][]{
        \minipage{1\linewidth}
        \lstset{style=lang,
        framerule=1pt,
        backgroundcolor=\color{white},
        rulecolor=\color{gray!50},
        frame=tblr,
        captionpos=b,
        #1}
}{\endminipage}

\newcommand{\techrep}[1]{#1}
\newcommand{\conference}[1]{}

\conference{\title{A Hoare Logic for Domain Specification}}
\techrep{\title{A Hoare Logic for Domain Specification (Full~Version)}}
\author{Eduard Kamburjan\inst{1} \and Dilian Gurov\inst{2}}
\institute{
  University of Oslo, Oslo, Norway, \email{eduard@ifi.uio.no}
\and
  KTH Royal Institute of Technology, Stockholm, Sweden, \email{dilian@kth.se}
}

\begin{document}

\maketitle

\begin{abstract}
Programs must be correct with respect to their application domain. Yet, the program specification and verification approaches so far only consider correctness in terms of computations.
In this work, we present a \emph{two-tier} Hoare Logic that integrates assertions for both implementation and domain.
For domain specification, we use description logics and semantic lifting, a recently proposed approach to interpret a program as a knowledge graph.
We present a calculus that uses translations between both kinds of assertions, thus separating the concerns in specification, but enabling the use of description logic in verification.
\end{abstract}

\section{Introduction}
Programs must respect constraints coming from their application domain, and thus, their correctness hast to rely on an encoding of domain knowledge. At the very minimum, application logic must correspond to business logic, but in extreme cases, such as simulators or applications in model-based engineering, the domain is directly encoded in the program. Description logics (DL) are an established tool to model domain knowledge with elaborate pragmatics in the form of, e.g., semantic web technologies; yet, making use of them for program specification and verification remains unexplored. 

In this work, we investigate reasoning about the correctness of programs  
with specification for both the \emph{implementation} (i.e., the program specifics) and its connection to the  application \emph{domain}. 
Domain-specific specification, in the form of description logic assertions, enables domain experts to be involved in modeling and programming, by giving them a tool to express their constraints without exposing them to implementation details. 
We aim to retain as much of the knowledge representation techniques and pragmatics during verification as possible, while making use of their logical foundation to recover assertions about the program:
Failed proof attempts should be interpreted and, for example, explained~\cite{DBLP:conf/exact/DengHS05,DBLP:conf/jelia/Schlobach04} in the domain. Similarly, keeping DL separate from program assertions enables the use of specialized solvers. Nonetheless, these assertions are used by a Hoare logic that operates only on the program state, and not on its interpretation in the domain.

\paragraph{Specification.}
To connect program state and description logics, we use ideas from \emph{semantically lifted programs}~\cite{DBLP:conf/esws/KamburjanKSJG21}. 
The state of a semantically lifted program is \emph{lifted} into the domain in the form of a \emph{knowledge graph}. This graph can then be enriched with DL axioms to interpret the program state in terms of the domain.

At the core of our approach are \emph{two-tier} specifications.  A two-tier assertion~$\driple{\liftphi}{\statephi}$ contains an assertion~$\statephi$ about the program state, and an assertion~$\liftphi$ about the domain, which specifies the \emph{lifted} state in terms of the domain. To connect the two assertions in the calculus, we lift not only the state, but also \emph{the specifications}, in order to 
recover information for $\statephi$ from $\liftphi$. 

\begin{figure}[t]
    \centering
    \includegraphics[width=100mm]{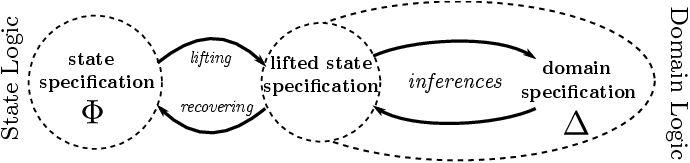}
    \caption{Relation between domain and state specifications in their respective logics.}
    \label{fig:relations}
\end{figure}

Fig.~\ref{fig:relations} illustrates the relations between state specification, the lifted state specification and the domain specification containing the lifted state specification. It is critical that the domain specification is using only the notions and vocabulary of the enriched state, and is not describing the lifted program state directly -- it is describing the lifted state \emph{enriched with additional axioms}. Thus, the program logic must be able to infer possible program states from the domain specification.

\paragraph{Verification.}
Consider a program that models the assembly of a car. Its domain ontology expresses concepts such as that a car $c$ has 4 wheels (\texttt{HasFourWheels}($c$)). Let us consider the following statement, that
sets the variable \texttt{wheels} to the parameter \texttt{nrWheels}. In the domain, the specification expresses that after execution, the modelled car has four wheels, i.e., is part of class \texttt{HasFourWheels}. For the implementation, it states that the parameter $\mathtt{nrWheels}$ must be 4.
\begin{align}
\driple{-}{\mathtt{nrWheels} \doteq 4} \mathtt{wheels := nrWheels} \driple{\mathtt{HasFourWheels}(c)}{-}\tag{1}\label{eq1}
\end{align}
From the perspective of the domain experts, the precondition cannot be stated, since they do not know how $c$ is modelled, and are not aware of the encoding of wheels as integers, i.e., of the very \emph{existence} of the variable \texttt{wheels}. Thus, both parts of the contract are stated from different perspectives and uphold the separation of concerns between domain and computation. But given a suitable specification lifting, we can transform the above two-tier triple into the following, and derive that to ensure the domain post-condition, the state must have set the variable \texttt{wheels} to 4. This is easily shown using a standard assignment rule.
\begin{align}
\driple{-}{\mathtt{nrWheels} \doteq 4} \mathtt{wheels := nrWheels} \driple{\mathtt{HasFourWheels}(c)}{\mathtt{wheels} \doteq 4}\tag{2}\label{eq2}
\end{align}
The remainder of this paper gives a precise description of the connection between state and domain specification required to set up a two-tier Hoare logic to enable such inferences. After introducing the needed preliminaries (Sec.~\ref{sec:prelim}), we make the above example more precise (Sec.~\ref{sec:example}), and give the used program logic (Sec.~\ref{sec:specify}) and its calculus (Sec.~\ref{sec:verify}). Finally, we give related work (Sec.~\ref{sec:related}) before we conclude.

\section{Preliminaries}\label{sec:prelim}
We give the basic definitions for the logic that we use to describe the states of the implemented program directly, as well as definitions for description logics for domain specification. To simplify terminology, we refer to the former as \emph{state logic} and to the later as \emph{domain logic}. 
Both logics are based on semantics defined over values $\mathsf{Val}$ that include data values, which in our case will be the integers~$\mathbb{Z}$ only, and names $\nabla$, which correspond to nominals in description logic.

To ease the later connection between the two logics, we split function symbols into functions that result in a name, and data functions that result in a data value, and consider the set of program variables in the signature.
\begin{definition}[Signatures]
A \emph{state signature} $\Sigma = \langle \mathtt{V}, \mathtt{F}, \mathtt{F}_d, \mathtt{P}\rangle$ is a tuple of variable names $\mathtt{V}$, function symbols $\mathtt{F}$, data function symbols $\mathtt{F}_d$, and predicate symbols $\mathtt{P}$.
A \emph{domain signature} $\Sigma_d = \langle \mathtt{N}, \mathtt{R}, \mathtt{T}, \mathtt{A}\rangle$ is a tuple of nominals $\mathtt{N}$, abstract roles $\mathtt{R}$, concrete roles $\mathtt{T}$, and atomic concepts $\mathtt{A}$.  
We say that a (state or domain) signature $\Sigma$ is a \emph{subsignature} of $\Sigma'$ ($\Sigma \subseteq \Sigma'$) if all its components are subsets. 
\end{definition}

In general, we refrain from treating arities formally and assume the usual framework to make formulas and interpretation respect the arity of function and predicate symbols.
\begin{definition}[Interpretations]
A \emph{state interpretation} $\mathcal{I}$ over a state signature $\Sigma= \langle \mathtt{V}, \mathtt{F}, \mathtt{F}_d, \mathtt{P}\rangle$ is
a map from:
\begin{enumerate*}
    \item function symbols $f \in \mathtt{F}$ to functions from values to names, 
    \item function symbols $f \in \mathtt{F}_d$ to functions from values to integers, and
    \item predicate symbols $p \in \mathtt{P}$ to functions from values to names.
\end{enumerate*}

A \emph{domain interpretation} $\mathcal{I}_d$ over a domain signature $\Sigma_d = \langle \mathtt{N}, \mathtt{R}, \mathtt{T}, \mathtt{A}\rangle$ is
a map from:
\begin{enumerate*}
    \item nominal symbols $o \in \mathtt{N}$ to names in $\nabla$, 
    \item abstract role symbols $R \in \mathtt{R}$ to relations over names,
    \item concrete role symbols $T \in \mathtt{T}$ to relations over names and $\mathbb{Z}$, and
    \item atomic concepts symbols $A \in \mathtt{A}$ to subsets of $\nabla$.
\end{enumerate*}

The set of all state interpretations is denoted $\mathbf{I}$, while the set of all domain interpretations is denoted $\mathbf{I}_d$.
\end{definition}
Program variables are not interpreted by $I$, but are part of the program state.

\begin{definition}[States and State Logic]\label{def:sl}
Let $\mathsf{V}$ be the set of program variables. 
A \emph{program state} $\state: \mathsf{V} \rightarrow \mathsf{Val}$ is a mapping from variables to values.
Let $\mathtt{S}$ denote the set of all program states.
Let $\Sigma= \langle \mathtt{V}, \mathtt{F}, \mathtt{F}_d, \mathtt{P}\rangle$ be a state signature.
State formulas $\statephi$ are defined by the following grammar, where $v$ ranges over $\mathtt{V}$, $p$ over $\mathtt{P}$, and $f$ over $\mathtt{F} \cup \mathtt{F}_d$. The set of all state formulas over $\Sigma$ is denoted $\Phi(\Sigma)$.
\[
\statephi ::= \statephi \wedge \statephi ~|~ \neg \statephi ~|~t \doteq t~|~ p(\overline{t}) 
\qquad\qquad t ::= v ~|~ x ~|~ f(\overline{t})
\]
\conference{The semantics of the state logic $\sigma, \mathcal{I} \models \Phi$ is defined relative to a program state and a state interpretation, and is given in the appendix.}
\techrep{The semantics of the state logic $\sigma, \mathcal{I} \models \Phi$ is defined relative to a program state and a state interpretation, and is given in Fig.~\ref{fig:fol:sem}.
\begin{figure}[b]
\begin{align*}
    \sigma, I &\models \statephi_1 \wedge \statephi_2 \iff \sigma, I \models \statephi_1 \text{ and } \sigma, I \models \statephi_2 \\
    \sigma, I &\models \neg\statephi \iff \sigma, I \not\models \statephi\\
    \sigma, I &\models t_! \doteq t_2 \iff \mathit{val}_{\sigma, I }(t_1) = \mathit{val}_{\sigma, I }(t_2)\\
    \sigma, I &\models p(t_1,\dots,t_n) \iff I(p)(\mathit{val}_{\sigma, I }(t_1),\dots, \mathit{val}_{\sigma, I }(t_n))\\ \\
    \mathit{val}_{\sigma, I }(v) &= \sigma(v) \qquad 
    \mathit{val}_{\sigma, I }(f(t_1,\dots,t_n)) = I(f)(\mathit{val}_{\sigma, I }(t_1),\dots, \mathit{val}_{\sigma, I }(t_n))
\end{align*}
     \caption{Semantics of the state logic.}
     \label{fig:fol:sem}
 \end{figure}}
\end{definition}

We use the usual abbreviations such as $\vee$ and~$\rightarrow$ and omit $\mathcal{I}$ in the satisfiability relation if it is understood.

We define a simple description logic, $\mathcal{ACLO}(\mathsf{D})$, following mostly the semantics of Horrocks and Sattler~\cite{DBLP:conf/ijcai/HorrocksS01} for $\mathcal{SHON}(\mathsf{D})$. We stress that our approach is not relying on any particular property of this logic (or any description logic), except for the presence of data types, and that we use it for the examples. We envision description logics as the most suited formalism for domain specification in our framework.

\begin{definition}[Description Logic]\label{def:dl}
Let $\Sigma_d = \langle \mathtt{N}, \mathtt{R}, \mathtt{T}, \mathtt{A}\rangle$ be a domain signature.
The syntax of domain formulas $\delta$ is defined by the following grammar, where $A$ ranges over $\mathtt{A}$,   $R$ over $\mathtt{R}$, $T$ over $\mathtt{T}$,  $o$ over $\mathtt{N}$, and $n$ over literals from $\mathbb{Z}$. The set of all domain formulas over $\Sigma_d$ is denoted $\Delta(\Sigma_d)$. We use $\liftphi$ to range over sets of domain formulas.
\begin{align*}
    \delta ::=\>& C \sqsubseteq C ~|~ C(o) ~|~ R(o,o) ~|~ R(o,n)\\
    C ::=\>& \top ~|~ \bot ~|~ A ~|~ \neg C ~|~ C \sqcup C ~|~ C \sqcap C ~|~ \exists R.\, C ~|~ \forall R.\, C ~|~ \exists T.\, n ~|~ \forall T.\, n
\end{align*}
\conference{The semantics $\mathcal{I}_d \models \delta$ is defined relative to a domain interpretation, and is given in the appendix.}
\techrep{The semantics $\mathcal{I}_d \models \delta$ is defined relative to a domain interpretation, and is given in Fig.~\ref{fig:dl:sem}.
\begin{figure}[t]
\begin{align*}
    \mathit{val}_\mathcal{I}(\top) &= \nabla \qquad\quad~  \mathit{val}_\mathcal{I}(\bot) = \emptyset \\
    \mathit{val}_\mathcal{I}(A) &= \mathcal{I}(A) \qquad \mathit{val}_\mathcal{I}(\neg C) = \nabla \setminus \mathit{val}_\mathcal{I}(C) \\
    \mathit{val}_\mathcal{I}(C \sqcup D) &= \mathit{val}_\mathcal{I}(C) \cup \mathit{val}_\mathcal{I}(D) \qquad
    \mathit{val}_\mathcal{I}(C \sqcap D) = \mathit{val}_\mathcal{I}(C) \cap \mathit{val}_\mathcal{I}(D) \\
    \mathit{val}_\mathcal{I}(\exists R.\, C) &= \{x \in \nabla ~|~ \exists y.~(x,y) \in \mathcal{I}(R) \wedge y \in \mathcal{I}(C)\} \\
    \mathit{val}_\mathcal{I}(\forall R.\, C) &= \{x \in \nabla ~|~ \forall y.~(x,y) \in \mathcal{I}(R) \wedge y \in \mathcal{I}(C)\}\\
    \mathit{val}_\mathcal{I}(\exists T.\, d) &= \{x \in \mathbb{Z} ~|~ \exists y.~(x,y) \in \mathcal{I}(T) \wedge y \in \mathcal{I}(T)\} \\
    \mathit{val}_\mathcal{I}(\forall T.\, d) &= \{x \in \mathbb{Z} ~|~ \forall y.~(x,y) \in \mathcal{I}(T) \wedge y \in \mathcal{I}(T)\}\\ 
    \mathcal{I} \models C \subseteq D &\iff \mathit{val}_\mathcal{I}(C) \subseteq \mathit{val}_\mathcal{I}(D)\qquad
    \mathcal{I} \models C(o) \iff \mathcal{I}(o) \in \mathit{val}_\mathcal{I}(C)\\
    \mathcal{I} \models R(o,o') &\iff (\mathcal{I}(c),\mathcal{I}(o') \in \mathcal{I}(R)\qquad
    \mathcal{I} \models T(o,d) \iff (\mathcal{I}(c),\mathcal{I}(d)) \in \mathcal{I}(T)
\end{align*}
    \caption{Semantics of the domain logic.}
    \label{fig:dl:sem}
\end{figure}}
We use the usual logic abbreviations such as~$\equiv$.
\end{definition}

Given a formula~$\Phi$, we denote the signature containing just the symbols it uses by $\mathsf{sig}~\Phi$ (resp.~$\mathsf{sig}~\Delta$).
Semantic entailment is defined as usual: Given two formulas $\Delta,\Delta'$, we say that $\Delta$ entails $\Delta'$, written $\Delta \models \Delta'$, if every interpretation that satisfies $\Delta$, also satisfies $\Delta'$. This naturally generalizes to sets of formulas. Given a set of domain formulas $\knowledge$, we write $\Delta \models^\knowledge \Delta'$ to denote that every interpretation that satisfies formula $\Delta$ and all elements of $\knowledge$ also satisfies $\Delta'$. 

\section{Motivating Example}\label{sec:example}
\paragraph{Scenario.} Consider a program that models the assembly of a small car, where a car is considered to be small if it has two doors and four wheels. This can be formalized in the domain logic using the following formula.
\[
\mathtt{SmallCar} \equiv \mathtt{HasTwoDoors} \sqcap \mathtt{HasFourWheels} \sqcap \mathtt{Car}
\]

Additionally, we know that everything that has a body is a car, and everything that has a chassis has a body. For doors, wheels, and the body of the car, we can formulate the following formulas to express that
everything that has 2 doors is part of the concept $\mathtt{HasTwoDoors}$, and analogously for $\mathtt{HasFourDoors}$ and $\mathtt{HasBody}$. We use the common pattern of \emph{stubs}~\cite{DBLP:conf/semweb/KrisnadhiH16a}: instead of modeling the number of doors using a \texttt{hasDoors} relation that maps to a number, we use a relation \texttt{doors} that maps to an individual that has some number associated with it using relation \texttt{hasValue}. As we see later, we can relate the stubs with variables in the programming language to connect the two formalisms.
\begin{align*}
\mathtt{HasBody} \sqsubseteq \mathtt{HasChassis} &\sqsubseteq \mathtt{Car}\qquad \exists \mathtt{doors}. \exists \mathtt{hasValue}. 2 \ensuremath{\equiv} \mathtt{HasTwoDoors}\\
\exists \mathtt{wheels}. \exists \mathtt{hasValue}. 4 &\ensuremath{\equiv} \mathtt{HasFourWheels}\qquad
\exists \mathtt{body}. \mathtt{NonZero} \ensuremath{\equiv}\mathtt{HasBody}\\
\neg\exists \mathtt{hasValue}.0 &\equiv \mathtt{NonZero}
\end{align*}

\begin{figure}[bt]
\begin{langcode}
var bodyId = 0; var wheels = 0; var doors = 0;
proc addWheels(nrWheels) begin 
  wheels := nrWheels; 
end;
proc assembly(id, nrDoors) begin 
  bodyId := id; addWheels(4); doors := nrDoors;
end 
\end{langcode}
    \caption{An assembly line program.}
    \label{fig:ex:prog}
\end{figure}

The program is given in Fig.~\ref{fig:ex:prog}. 
The assembly is old-fashioned: it starts with a chassis, and has three substeps, namely adding the body, by assigning a non-zero id, then adding the wheels (\texttt{addWheels}), and adding the doors. It is operating on a single car, which is modelled by the variable \texttt{bodyId} for the id of the body, where $\mathtt{bodyId} = 0$ models that no body is attached, the variable \texttt{doors} which models the number of doors on the body, and \texttt{wheels}, which models the number of wheels. The considered car has a chassis, which is not explicit in the program.

\paragraph{Specification.} Our aim is to specify that procedure \texttt{assembly} indeed assembles a small car. However, the domain expert has no knowledge about the computational encoding of the process, e.g., that the wheels are modelled as a global variable. 
The contract of procedure \texttt{assembly} is as follows. In the beginning we get the number of doors (which must be 2) and the id of the body (which must be non-null), and in the end it is a small car.
The individual $c$ is implicit in the program -- in our example, the program assembles exactly one car, but this information is not relevant for the domain expert. As specifications, we use pairs $
\driple{\liftphi}{\statephi}$ that express that domain formula $\liftphi$ and state formula $\statephi$ must hold. 
\[
\driple{-}{\mathtt{nrDoors} \doteq 4 \wedge \mathtt{bodyId} \neq 0}\mathtt{assembly()}\driple{\mathtt{SmallCar}(c)}{-}
\]

\noindent Let us now turn to the specification of $\mathtt{addWheels}$.
The domain specification explains what is expected from the view of the car assembly (the car already has a chassis), while the implementation specification ($\mathtt{nrWheels} = 4$) specifies additional conditions \emph{not visible in the domain} to ensure correctness. The former is specified by the domain expert, while the latter is added by the programmer.

\[
\driple{-}{\mathtt{nrWheels} \doteq 4} \mathtt{addWheels(nrWheels)} \driple{\mathtt{HasFourWheels}(c)}{-}
\]
The post-condition is obvious - it states that afterwards the car being assembled is part of class \texttt{HasFourWheels}. Its domain precondition states that the car has a chassis before.
Note that the implementation details are hidden from the domain experts -- they do not know how $c$ is modelled, whether it always has a chassis in the program, or whether this is explicit. They are, thus, not able to state the state precondition, as they are not aware of the encoding of wheels. Thus, the two parts of the contracts are stated from different perspectives and uphold the \emph{separation of concerns} between domain and computation.
Furthermore, we stress that the specification at the level of procedure contracts enables the domain expert a more fine-grained specification, without being exposed to many technicalities, but requires that we must be able to switch between a domain and a state view in the middle of the analyzed statement.

\paragraph{Verification.}
To verify that $\mathtt{addWheels}$ adheres to its specification, we have to show that its procedure body indeed transforms a car into one with four wheels, which is exactly Eq.~\ref{eq1}.

In a classical weakest precondition calculus, we would now substitute $\mathtt{wheels}$ by $\mathtt{nrWheels}$ in the post-condition -- the post-condition obviously needs to be $\mathtt{wheels} \doteq 4$. But in out setting we only have the domain specification. Instead of introducing redundancy in the specification, which would also break our separation between tasks for the domain expert and tasks for the programmer, we can retrieve a state post-condition as follows.

At its basis, we rely on semantic lifting, which generates a domain state from a program state.
Let us consider the program state $\sigma_0$ with $\sigma_0(\mathtt{wheels}) = 4$. Its lifting consists of axioms for the program state, information about the domain \emph{and} additional formulas that connect the domain concepts with those describing the lifted program state. Those are given in Fig.~\ref{fig:ex:lift}.
Note that the resulting knowledge graph has two parts: lifted program state, and domain knowledge. However, the domain specification is only concerned with the domain knowledge. The first part is generic for the program, e.g., the existence of variables -- instead of designing a new lifting for every application, this \emph{direct lifting} can be used as a basis to simplify modeling~\cite{DBLP:conf/esws/KamburjanKSJG21}. 

\begin{figure}[t]
\begin{align*}
    &\mathtt{hasValue}(\mathtt{wheelsVar}, 4) \qquad 
\mathtt{HasChassis}(c) \qquad \mathtt{wheels}(c, \mathtt{wheelsVar}) \\
&\mathtt{body}(c, \mathtt{bodyVar}) \qquad\mathtt{doors}(c, \mathtt{doorsVar})\\[-7mm]
\end{align*}
    \caption{The first formula is (part of) the lifted state, the last four formulas connect lifted program state and domain knowledge. }
    \label{fig:ex:lift}
\end{figure}

Still, we can deduce knowledge about the lifted state: 
If the car has four wheels  ($\mathtt{HasFourWheels}(c)$), then the corresponding variable must be set to four ($\mathtt{hasValue(wheels,4)}$).
This information, in turn, can be interpreted in the program logic as $\mathtt{wheels} \doteq 4$, in order to strengthen our specification into Eq.~\ref{eq2}.

Using the rule for assignment, we can prove the correctness of $\mathtt{addWheels}$ w.r.t.\ to its specification. We must consider the relation of $\mathtt{HasFourWheels}(c)$ and $\mathtt{wheels} \doteq 4$ -- as the program must establish both conditions, but only controls the state post-conditions, the state post-condition $\mathtt{wheels} \doteq 4$ must imply the complete domain post-condition $\mathtt{HasFourWheels}(c)$. 
Having established our example and illustrated the challenges therein. we now give a formal treatment of the underlying Hoare logic. We return to the assembly line after presenting the calculus in Sec.~\ref{sec:verify}, and show that the whole program is indeed correct.

\section{A Two-Tier Hoare Logic}\label{sec:specify}
Our task is now to ensure that the program indeed models the assembly of a small car at the domain level, not just through the name of its variables and procedures. 
Our approach is based on two-tier assertions: A two-tier assertion has two parts, or tiers, in different logics, that are connected through a lifting mechanism for translation.
To do so, we must first define
how to interpret a state in the domain and define the semantic lifting of a state.

\begin{definition}[State and Specification Lifting]
A \emph{state lifting} is defined as a function $\liftstate: \mathtt{S}\times \mathbf{I} \rightarrow \mathbf{I}^d$ from program models to domain models.

A \emph{specification lifting} $\liftspec: \statephi(\Sigma) \rightarrow \liftphi(\Sigma^d)$ is a mapping from program formulas to domain formulas. We denote the signature of the images of $\liftspec$ as its \emph{kernel}, written 
\(
\mathbf{ker}~\liftspec = \bigcup_{\Phi \in \statephi(\Sigma)}\mathsf{sig}\big(\liftspec(\Phi)\big)
\).
\end{definition}
State and specification lifting must be compatible, in the sense that if a state satisfies a state formula, then its lifting must satisfy the lifted assertion. This is required to argue about the soundness of the specification lift $\liftspec$ -- the state lifting $\liftstate$ is not used in the calculus we give later.
\begin{definition}[Compatibility]\label{def:compatible}
A pair $(\liftstate,\liftspec)$ is \emph{compatible} w.r.t. a state interpretation $\mathcal{I}$ and a set of domain formulas $\knowledge$ iff lifting state and formula preserves satisfaction:
    $\forall \state \in \mathtt{S}.\, (\state, \mathcal{I} \models \statephi \,\Rightarrow\, \liftstate(\state) \models_\knowledge \liftspec(\statephi))$.
\end{definition}

The domain logic is less expressive then the state logic. Its task is to provide a way to (model and) specify the domain without exposing implementation details. Applying $\liftspec$ allows one to interpret an intermediate state specification in the domain, for example to examine what this intermediate state is modelling. Similarly, the domain specification is not only part of the pre- and post-condition of the program, but also part of the pre- and post-condition of \emph{procedures}. The lifting $\liftspec$ is, thus, also needed to add information to apply these contracts.

However, the program itself is analyzed in terms of the state logic. For example, the effect of an assignment can be clearly expressed for the state specification, but not for the domain. Here, we require to \emph{recover} information from the domain specification by applying the inverse $\liftspec^{-1}$.

Consider the state specification $\phi = \mathtt{wheels}\doteq 4$ and the domain specification $\delta = \mathtt{HasFourWheels}(c)$. The lifting $\liftspec$ enables us, together with further inferences and assuming a fitting pair of liftings, to derive $\delta$ from $\phi$, and the recovering mapping $\liftspec^{-1}$ enables us to derive $\phi$ form $\delta$. Before we connect state and domain specification further, we 
give a direct lifting.



The characteristic formula $\chi_\sigma$ of a state $\sigma$ is defined as 
\(
\bigwedge_{v\in \mathbf{dom}\sigma} v \doteq \sigma(v)
\).

\begin{definition}[Direct Lifting]
The specification lifting $\liftspec_\mathsf{direct}$ is defined by:
\begin{align*}
    \liftdirect(v~\doteq~n) &= \{\mathsf{hasValue}(\mathtt{var}_v, n)\} \qquad
    \liftdirect(v~\neq~0) = \{\mathtt{NonZero}(\mathtt{var}_v)\} \\
    \liftspec_\mathsf{direct}(\statephi_1 \wedge \statephi_2) &= \liftspec_\mathsf{direct}(\statephi_1) \cup \liftdirect(\statephi_2)  
\end{align*}
The state lifting is defined by:
\(
\liftstate_\mathsf{direct} = \mathcal{I} \text{ such that } \mathcal{I} \models\liftdirect(\chi_\state)
\).
\end{definition}

The pair $(\liftstate_\mathsf{direct}, \liftspec_\mathsf{direct})$ is compatible, and the example in Fig.~\ref{fig:ex:lift} is an application of it with $\knowledge = \{\mathtt{HasChassis}(c),\mathtt{wheels}(c,\mathtt{wheelsVar}),\dots\}$. The variables are also modelled as stubs -- the formula $\mathtt{wheels}(c,\mathtt{wheelsVar})$ indeed expresses that the variable $\mathtt{wheelsVar}$ (i.e., $\mathtt{var}_\mathtt{wheels}$) is the stub that can be used in the domain to reason about the wheels of the car $c$. The kernel of $\liftspec_\mathsf{direct}$ is 
\(
\mathbf{ker}~\liftspec_\mathsf{direct} = \big\{\mathtt{hasValue},\mathtt{NonZero}\big\} \cup \big\{\mathtt{var}_v ~|~ v \in \mathbf{dom}~\sigma\big\} \cup \mathsf{sig}(\knowledge)
\). Note the explicit addition of $\mathtt{NonZero}$, which enables us to lift (and recover from) more abstract specifications than characteristic formulas.

\subsection{Assertions}
Equipped with a formal definition of lifting, we can now define specifications that have both a domain and a state component. We refer to such specifications as \emph{two-tier assertions}.
\begin{definition}[Two-Tier  Assertion]
\label{def:liftedasrt}
Let  $(\liftstate,\liftspec)$ be a compatible set of liftings (w.r.t. some $\mathcal{I}$ and $\knowledge$).
Let $\Delta$ range over sets of domain formulas over $\Sigma_d$ and $\statephi$ over state formulas over $\Sigma$. A \emph{two-tier assertion} has the form 
\[
\driple{\liftphi}{\statephi}
\]
written $\sdriple{\liftphi}{\statephi}$ for brevity, and has the following semantics 
\[
\sigma \models_\knowledge \driple{\liftphi}{\statephi} \quad\mathit{iff}\quad \sigma, \mathcal{I} \models \statephi \text{ and } \liftstate(\sigma), \liftspec(\statephi) \models^\knowledge  \liftphi
\]
We say that a two-tier assertion 
is \emph{strongly consistent} if $\liftspec(\statephi) \models_\knowledge \liftphi$.
\end{definition}

In a strongly consistent assertion, the domain is determined entirely by the state, which is exactly the condition we discussed above for post-conditions.

A program, which we define below, is specified using contracts.
A contract is one of the places where domain specification can be used -- we do not expect the domain expert to annotate intermediate specification in sequences of statements, but to interact with the developer on the level of procedures and eventually other, abstracting language constructs.
 \begin{definition}[Procedure Contract]\label{def:contract}
 A \emph{contract} for a procedure $\mathtt{p}(\mathtt{v})$ is a pair of two-tier assertions $\left(\driple{\liftphi_1^\mathtt{p}}{\statephi_1^\mathtt{p}},\driple{\liftphi_2^\mathtt{p}}{\statephi_2^\mathtt{p}}\right)$, called the \emph{precondition} and the \emph{postcondition}, respectively.
 The set of all contracts in a program is denoted $\contracts$.
 Retrieving the precondition (resp.\ postcondition) of a procedure~$\mathtt{p}$ with parameter~$\mathtt{e}$ replacing variable~$\mathtt{v}$ is denoted as follows. 
    \[ \mathtt{Pre}_\contracts(\mathtt{p},\mathtt{e}) = \driple{\liftphi_1^\mathtt{p}}{\statephi_1^\mathtt{p}[\mathtt{v}\setminus \mathtt{e}]} \qquad \mathtt{Post}_\contracts(\mathtt{p},\mathtt{e}) = \driple{\liftphi_2^\mathtt{p}}{\statephi_2^\mathtt{p}[\mathtt{v}\setminus \mathtt{e}]}\]
 \end{definition}


We can now introduce a simple, imperative programming language with procedure calls that operates on states. 
As we are not concerned with expressive power here, we limit expressions to a minimum, and procedures to only one parameter. The semantics of the language is relative to a set of contracts. This simplifies the later definition; a non-relative version is easily obtained by inlining.

 \begin{definition}[Programming Language]\label{def:pl}
 The \emph{syntax} of our programming language is defined by the following grammar.
 Let $\mathtt{v}$ range over variables, $n$ over literals, and $\mathtt{p}$ over procedure names.
 \begin{align*}
     \mathsf{prog} ::=&\many{\mathbf{var}~\mathtt{v}~=~\mathsf{expr}~;}~\many{\mathsf{proc}} && \text{programs}\\
     \mathsf{proc} ::=& \mathtt{p}(\mathtt{v})\ \mathbf{begin}~\statement~\mathbf{end} \qquad \mathsf{expr} ::= n ~|~\mathtt{v} && \text{procedures and expressions}\\
     \statement ::=& \mathtt{v}~:=~\mathsf{expr}; ~|~ \statement\, ;\, \statement &&\\~
     & |~ \mathbf{if}\ (\mathsf{expr})\ \mathbf{then}~\statement~\mathbf{else}~\statement~\mathbf{fi} &&\\~
     & |~ \mathbf{while}\ (\mathsf{expr})\ \mathbf{do}~\statement~\mathbf{od} ~|~ \mathtt{p}(\mathsf{expr}) && \text{statements}
 \end{align*}

 The \emph{semantics} of our programming language is defined relative to a contract set~$\contracts$ and a set of formulas~$\knowledge$, as a binary relation on states, i.e., as a \emph{denotational semantics} $\llbracket \statement \rrbracket \subseteq \states \times \states$, shown in Fig.~\ref{fig:progsem}, where we use $\mathsf{LFP } F$ to denote the least fixed-point of a function~$F$. 
\begin{figure}[t]
\[\begin{array}{rcl}
   \mathsf{cond} (P, R_1, R_2)  & \>=\> & \setdef{(\sigma, \sigma')}{(\sigma \in P \wedge (\sigma, \sigma') \in R_1) \vee (\sigma \not\in P \wedge (\sigma, \sigma') \in R_2)} \\
   F_{\contracts, \knowledge} (R) & \>=\> & \mathsf{cond} (\bexdenot{\mathsf{expr}}, \denotCK{\statement} \circ R, \mathsf{id}_\states) \\ \\
   \denotCK{\mathtt{v}~\mathtt{:=}~\mathsf{expr}} & \>=\> & \setdef{(\state, \state')}{\state' = \state [\mathtt{v} \mapsto \arexdenot{\mathsf{expr}}]} \\
   \denotCK{\statement_1\, ;\, \statement_2}  & \>=\> & \denotCK{\statement_1} \circ \denotCK{\statement_2} \\
   \denotCK{\mathbf{if}\ (\mathsf{expr})\ \mathbf{then}~\statement_1~\mathbf{else}~\statement_2~\mathbf{fi}} & \>=\> & \mathsf{cond} (\bexdenot{\mathsf{expr}}\!, \denotCK{\statement_1}\!, \denotCK{\statement_2}) \\
   \denotCK{\mathbf{while}\ (\mathsf{expr})\ \mathbf{do}~\statement~\mathbf{od}} & \>=\> & \mathsf{LFP}\ F_{\contracts, \knowledge} \\
   \denotCK{\mathtt{p}(\mathsf{expr})}  & \>=\> &  \left\{(\sigma, \sigma') ~|~ \sigma \models_\knowledge \mathtt{Pre}_\contracts(\mathtt{p},\arexdenot{\mathsf{expr}}_\sigma) \wedge \sigma' \models_\knowledge \mathtt{Post}_\contracts(\mathtt{p},\arexdenot{\mathsf{expr}})\right\} \\
\end{array} \]
    \caption{Program semantics.}
    \label{fig:progsem}
\end{figure}
\end{definition}
Our semantics is \emph{procedure-modular}, i.e., we define the semantics of a procedure call as the semantics of the contract of the called procedure. This is why our semantics is based on binary relations on states rather than on partial functions.
The evaluation functions $\arexdenot{\cdot}$ and $\bexdenot{\cdot}$ of arithmetic and boolean expressions are standard and omitted for brevity. 
For an extended treatment of the used kind of denotational semantics we refer to the standard texts~\cite{win-93-book,nie-nie-07-book}, and for details about the treatment of contracts to~\cite{gur-wes-18}.

\subsection{Hoare Triples}
A two-tier assertion specifies a state, while two-tier Hoare triples relate the initial and final states of a program execution: If the precondition holds in the initial state, and the program terminates, then the postcondition holds in the final state. In our case, the pre- and post-conditions are lifted assertions.
\begin{definition}[Two-Tier Hoare Triple]
\label{def:lifted-hoare-triple}
A \emph{two-tier Hoare triple} with respect to a compatible mapping $(\liftstate,\liftspec)$ has the following form
\[
\driple{\liftphi_1}{\statephi_1} \statement \driple{\liftphi_2}{\statephi_2}
\]
with the expected semantics (given sets of formulas $\knowledge$ and contracts~$\contracts$) 

\noindent\scalebox{0.95}{
\parbox{1cm}{
\begin{align*}
&(\sigma, \sigma') \models_\knowledge \driple{\liftphi_1}{\statephi_1} s \driple{\liftphi_2}{\statephi_2}\quad\textit{iff.}\quad(\sigma, \sigma') \in \denotCK{s} \wedge 
\left(\state \models_\knowledge \driple{\liftphi_1}{\statephi_1}
\rightarrow \state' \models_\knowledge \driple{\liftphi_2}{\statephi_2}\right)
\end{align*}
}}
\end{definition}


Let us now turn to the recovering mapping $\liftspec^{-1}$. 
This faces the challenge of ``delifting'' arbitrarily formulas, while $\liftspec$ must merely lift a limited set of expressions, on which we can easily enforce a normal form.
Recovering could only operate on the limited signature $\mathbf{ker}~\liftspec$. 
For this reason, we must be able to infer formulas from a domain specification that are within this limited signature. While in some cases we may be able to deduce them, in the general case we may have to rely on \emph{abduction}. 

Fortunately, abduction is feasible in our setup -- we have a clear notion of abductibles through the signature, and we only require formulas about individuals, not arbitrary formulas. This is exactly the well-explored setting of ABox abduction with abductibles~\cite{pat}.

We abstract from the exact mechanism to generate the inversible kernel and assume just some function that realizes it.
\begin{definition}[Kernel-Generator]\label{def:abduct}
Let $\liftspec$ be a specification lifting.
A \emph{kernel-generator} is a function $\abduct_\knowledge:2^\Delta \rightarrow 2^\Delta$ that, given a set of domain formulas~$\Delta$, generates another set of domain formulas~$\abduct_\knowledge(\Delta)$ such that 
$\mathsf{sig}\big(\abduct_\knowledge(\Delta)\big) \subseteq \mathbf{ker}~\liftspec$ and $\abduct_\knowledge(\Delta) \models_\knowledge \Delta$. 
\end{definition}

A kernel-generator can either perform abduction or deduction.
In the case of abduction, most formulas are essentially implications that have  $\mathbf{ker}~\liftspec$ as the consequent and the rest of the signature in the antecedent (e.g., \textit{``if the program variable has this value, then the domain individual belongs to this concept''}). Abduction is not sound, but may still be useful to generate a condition for program correctness.\footnote{One can precisely specify abduction with the usual conditions on $\abduct_\knowledge$~\cite{DBLP:journals/igpl/MayerP93}.} In case of deduction, the kernel-generator infers the only possible values for the variables.

Given a state specification, we can lift it using $\liftspec$ to a set of formulas with signature $\mathbf{ker}~\liftspec$. From there, we can deduce further formulas about the domain using description logic reasoning.
To compare a given set of domain formulas, we use the kernel generator $\alpha$ to get formulas with signature $\mathbf{ker}~\liftspec$, from which we can inverse the lifting $\liftspec^{-1}$.
Implication on assertions is lifted as expected.
\begin{definition}\label{def:liftimpl} One two-tier assertion implies another if the following holds.
    \[\driple{\liftphi}{\statephi} \rightarrow_\knowledge \driple{\liftphi'}{\statephi'} \quad\mathit{iff.}\quad \forall \sigma. \left (\sigma \models_\knowledge \driple{\liftphi}{\statephi} \rightarrow \sigma\models_\knowledge \driple{\liftphi'}{\statephi'} \right ) \]
\end{definition}


Formally, we can now express the relations between $\liftspec$, $\alpha$ and $\liftspec^{-1}$ with the following lemma, on which our calculus will heavily rely.
\begin{lemma}\label{lem:conditions}
The following three implications and equivalences hold.
\begin{enumerate}
    \item 
Generating the kernel of a domain specification implies the original domain specification:
\(
    \sdriple{\liftphi,\abduct_\knowledge(\liftphi)}{\statephi} \rightarrow_\knowledge  \sdriple{\liftphi}{\statephi} 
\).
\item Adding lifted specification preserves satisfiability:
\(
    \sdriple{\liftphi}{\statephi} \leftrightarrow_\knowledge  \sdriple{\liftphi,\liftspec(\statephi)}{\statephi}. 
\)
\item Adding recovered specification preserves satisfiability:
\[
    \driple{\liftphi,\liftphi'}{\statephi} \leftrightarrow_\knowledge  \driple{\liftphi,\liftphi'}{\statephi \wedge \liftspec^{-1}(\liftphi')}  \qquad\text{ where }\mathsf{sig}(\liftphi') \subseteq \mathbf{ker}~\liftspec
\]
\end{enumerate}
\end{lemma}
\techrep{
\begin{proof}
 Property $1.$ holds because $\alpha(\Delta)$ is a weakening and does not impose additional conditions. From Def.~\ref{def:abduct} we have
    \[\alpha_\knowledge(\Delta)\models_\knowledge \Delta \iff \forall \mathcal{I}.~\mathcal{I} \models_\knowledge\alpha_\knowledge(\Delta)\rightarrow \mathcal{I}\models_\knowledge \Delta\]
    So, the following relation holds by the semantics of the domain logic. \[\mathcal{I} \models_\knowledge\alpha_\knowledge(\Delta), \Delta~\rightarrow~ 
    \mathcal{I} \models_\knowledge \Delta\]
    This is trivially applied to lifted assertions, where $\mathcal{I}$ can be constructed from~$\sigma$.
    \[
    \sigma \models_\knowledge \driple{\liftphi,\alpha_\knowledge(\Delta)}{\Phi} \rightarrow 
    \sigma \models_\knowledge \driple{\liftphi}{\Phi} 
    \]
Properties $2.$ and $3.$ follow directly from Def.~\ref{def:compatible} and Def.~\ref{def:liftedasrt}.
\end{proof}}

\section{A Calculus for a Two-Tier Hoare Logic}\label{sec:verify}
The calculus combines the concepts introduced in the previous section by integrating two systems of rules: The first implements a weakest-precondition calculus on the implementation-specification for each statement, except for procedure calls. These rules erase domain information, as every change in the implementation can effect any formula in the lifted specification.
The second system of rules implements the kernel-generation, lifting and recovering that enables us to restore this information, or to add information to the implementation from the domain before it is erased.

 Verification is compositional, and always local to a single procedure and relative to the context, i.e., the contracts and used background knowledge.
 Our judgement is, thus, verifying a lifted Hoare triple in a fixed context.
 \begin{definition}[Calculus]
 \label{def:judgement}
 Let $\contracts$ be the set of contracts for a given program, and $\knowledge$ a set of formulas.
 A \emph{judgement} of the calculus has the following form.
    \[ \contracts, \knowledge \vdash \driple{\liftphi_1}{\statephi_1} \statement \driple{\liftphi_2}{\statephi_2}\]
    We say that the judgement is \emph{valid}, if for every state, in every terminating run where all procedures adhere to their respective contract, the Hoare triple holds.
\begin{align*}
&\contracts, \knowledge \models \driple{\liftphi_1}{\statephi_1} \statement \driple{\liftphi_2}{\statephi_2} \enskip\mathit{iff}\enskip  \forall (\sigma, \sigma') \in \denotCK{s}.
\left(\state \models_\knowledge \driple{\liftphi_1}{\statephi_1}
\rightarrow \state' \models_\knowledge \driple{\liftphi_2}{\statephi_2}\right)
\end{align*}
    
\noindent Let $P_1, \dots, P_n$ and $C$ be judgements. A \emph{rule} has the following form.
\begin{prooftree}
\AxiomC{$P_1 \dots P_n$}
\UnaryInfC{$C$}
\end{prooftree}
The rule is termed \emph{sound}, if validity of the premises $P_1,\dots,P_n$ implies validity of the conclusion~$C$.
\end{definition}

To connect the two specifications we require a set of rules to modify the lifted assertions that serve as pre- and postconditions, as well as to strengthen the precondition or weaken the postcondition. These rules, given in Fig.~\ref{fig:calc2}, are all given relative to some $\knowledge,\abduct_\knowledge,\mathcal{I}$ and a compatible pair $(\liftstate,\liftspec)$, and implement the connection between domain and computation specification. 
In detail, rules \rulename{pre-lift} and \rulename{post-lift} enable to lift the state specification. 
Rules \rulename{pre-core} and \rulename{post-core} abduct a core in the domain specification. We remind here that we define~$\abduct$ so that the signature of its
range indeed is the kernel. 
Rules \rulename{pre-inv} and \rulename{post-inv} apply the inverse lifting on the core. 
The consequence rule enables to strengthen, respectively weaken, the specification. 
We stress here that \rulename{pre-core} and \rulename{post-core} (and \rulename{var}, see below) invoke a DL reasoner -- keeping $\statephi$ and $\liftphi$ separate enables us to do so, an approach which merely translate DL into first-order logic would require to pass the verification condition to a solver for less tractable logics.

\begin{figure}[bt]
\begin{minipage}{.5\textwidth}
\begin{prooftree}
\AxiomC{$\contracts,\knowledge \vdash \driple{\liftphi_1,\liftspec(\statephi_1)}{\statephi_1} \statement \driple{\liftphi_2}{\statephi_2}$}
\LeftLabel{\rulename{pre-lift}}
\UnaryInfC{$\contracts,\knowledge \vdash \driple{\liftphi_1}{\statephi_1} \statement \driple{\liftphi_2}{\statephi_2}$}
\end{prooftree}
\begin{prooftree}
\AxiomC{$\liftphi_1 \models^\knowledge \abduct_\knowledge(\liftphi_1)$}
\noLine
\UnaryInfC{$\contracts,\knowledge \vdash \driple{\liftphi_1,\abduct_\knowledge(\liftphi_1)}{\statephi_1} \statement \driple{\liftphi_2}{\statephi_2}$}
\LeftLabel{\rulename{pre-core}}
\UnaryInfC{$\contracts,\knowledge \vdash \driple{\liftphi_1}{\statephi_1} \statement \driple{\liftphi_2}{\statephi_2}$}
\end{prooftree}
\end{minipage}
\begin{minipage}{.5\textwidth}
\begin{prooftree}
\AxiomC{$\contracts,\knowledge \vdash \driple{\liftphi_1}{\statephi_1} \statement \driple{\liftphi_2,\liftspec(\statephi_2)}{\statephi_2}$}
\LeftLabel{\rulename{post-lift}}
\UnaryInfC{$\contracts,\knowledge \vdash \driple{\liftphi_1}{\statephi_1} \statement \driple{\liftphi_2}{\statephi_2}$}
\end{prooftree}
\begin{prooftree}
\AxiomC{$\liftphi_2 \models^\knowledge \abduct_\knowledge(\liftphi_2)$}
\noLine
\UnaryInfC{$\contracts,\knowledge \vdash \driple{\liftphi_1}{\statephi_1} \statement \driple{\liftphi_2,\abduct_\knowledge(\liftphi_2)}{\statephi_2}$}
\LeftLabel{\rulename{post-core}}
\UnaryInfC{$\contracts,\knowledge \vdash \driple{\liftphi_1}{\statephi_1} \statement \driple{\liftphi_2}{\statephi_2}$}
\end{prooftree}
\end{minipage}

\begin{prooftree}
\AxiomC{$\contracts,\knowledge \vdash \driple{\liftphi_1}{\statephi_1} \statement \driple{\liftphi,\liftphi_2}{\statephi_2\wedge\liftspec^{-1}(\liftphi_2)}$}
\LeftLabel{\rulename{post-inv}}
\RightLabel{$\mathsf{sig}(\liftphi_2) \subseteq \mathbf{ker}~\liftspec$}
\UnaryInfC{$\contracts,\knowledge \vdash \driple{\liftphi_1}{\statephi_1} \statement \driple{\liftphi,\liftphi_2}{\statephi_2}$}
\end{prooftree}
\begin{prooftree}
\AxiomC{$\contracts,\knowledge \vdash \driple{\liftphi,\liftphi_1}{\statephi_1\wedge\liftspec^{-1}(\liftphi_1)} \statement \driple{\liftphi_2}{\statephi_2}$}
\LeftLabel{\rulename{pre-inv}}
\RightLabel{$\mathsf{sig}(\liftphi_1) \subseteq \mathbf{ker}~\liftspec$}
\UnaryInfC{$\contracts,\knowledge \vdash \driple{\liftphi,\liftphi_1}{\statephi_1} \statement \driple{\liftphi_2}{\statephi_2}$}
\end{prooftree}

\begin{prooftree}
\AxiomC{$\driple{\liftphi_1}{\statephi_1} \rightarrow_\knowledge \driple{\liftphi_1'}{\statephi_1'}$}
\AxiomC{$\contracts,\knowledge \vdash \driple{\liftphi_1'}{\statephi_1'} \statement \driple{\liftphi_2'}{\statephi_2'}$}
\AxiomC{$\driple{\liftphi_2'}{\statephi_2'} \rightarrow_\knowledge \driple{\liftphi_2}{\statephi_2}$}
\LeftLabel{\rulename{cons}}
\TrinaryInfC{$\contracts,\knowledge \vdash \driple{\liftphi_1}{\statephi_1} \statement \driple{\liftphi_2}{\statephi_2}$}
\end{prooftree}
    \caption{Rules for manipulating pre- and post-conditions.}
    \label{fig:calc2}
\end{figure}

The rules for statements are given in Fig.~\ref{fig:calc1}. 
Using the previously introduced rules we can easily derive more complex rules that operate on both levels.

Rule \rulename{var} is the assignment rule for variables. On the state level, it is exactly the rule from the original Hoare calculus, 
expressing the precondition as the syntactically updated postcondition. On the domain level, it expresses that any domain knowledge in the domain postcondition must be justified by the 
state post-condition. As the domain precondition, however, it erases all information as the assignment may have arbitrary effects on the domain. Note that we can erase the domain knowledge in practice -- strong consistency does not imply equivalence. 
In detail, \rulename{skip} expresses that the \texttt{skip} statement has no effect on the state.
Branching, handled by rule \rulename{branch}, also erases the domain precondition, as it modifies the state precondition.
Rule \rulename{inv} handles loops by unrolling.
Lastly, rule \rulename{contract} just checks that the contract is adhered to, and \rulename{seq} is as expected completely analogous to the original rule.
It is worth noting that rule \rulename{contract} uses domain specification only to syntactically match it with the $\mathsf{Pre}$ and $\mathsf{Post}$ predicates.
Our main result is the soundness of the rules for lifted Hoare triples.
\begin{theorem}[Soundness]\label{thm:sound}
The rules in Fig.~\ref{fig:calc2} and Fig.~\ref{fig:calc1} are sound.
\end{theorem}

\techrep{
\begin{proof}
    The soundness of rules \rulename{pre-lift} and \rulename{post-lift} follows directly from property~$2.$ of Lemma~\ref{lem:conditions}. Similarly, the soundness of rules \rulename{pre-inv} and \rulename{post-inv} follows directly from property~$3.$ of Lemma~\ref{lem:conditions}. Finally, rules \rulename{pre-core} and \rulename{post-core} are a direct consequence of property~$1.$ of Lemma~\ref{lem:conditions}. 
    Rule \rulename{cons} is not specific to our calculus and is the standard consequence rule on a semantic level.

    Let us turn to the rules for statements. 
    \begin{itemize}
        \item Rule \rulename{skip}. 
We have 
\begin{align*}
    &\forall (\sigma,\sigma') \in \llbracket \mathbf{skip} \rrbracket_{\contracts,\knowledge}.~\left(\sigma \models_\knowledge \driple{\liftphi}{\statephi} \rightarrow \sigma' \models_\knowledge \driple{\liftphi}{\statephi} \right)\\
    \iff&\forall \sigma.~\left(\sigma \models_\knowledge \driple{\liftphi}{\statephi} \rightarrow \sigma \models_\knowledge \driple{\liftphi}{\statephi} \right) \iff \mathbf{true}
\end{align*}
        \item Rule \rulename{seq}. 

We have, by validity of the premises
\begin{align}
    &\forall (\sigma,\sigma'') \in \llbracket \statement_1 \rrbracket_{\contracts,\knowledge}.~\left(\sigma \models_\knowledge \driple{\liftphi_1}{\statephi_1} \rightarrow \sigma'' \models_\knowledge \driple{\liftphi_3}{\statephi_3} \right)\\
    &\forall (\sigma'',\sigma') \in \llbracket \statement_2\rrbracket_{\contracts,\knowledge}.~\left(\sigma'' \models_\knowledge \driple{\liftphi_3}{\statephi_3} \rightarrow \sigma' \models_\knowledge \driple{\liftphi_2}{\statephi_2} \right)
\end{align}
Additionally, by the definition of~$\circ$, we have
\[
\forall \sigma,\sigma'.~(\sigma,\sigma') \in \llbracket \statement_1;\statement_2 \rrbracket_{\contracts,\knowledge} \rightarrow
\exists \sigma''.~(\sigma,\sigma'') \in \llbracket \statement_1 \rrbracket_{\contracts,\knowledge}  \wedge (\sigma'',\sigma') \in \llbracket \statement_2 \rrbracket_{\contracts,\knowledge} 
\]
By applying the first premise we get that if $\sigma$ is a model for $\sdriple{\liftphi_1}{\statephi_1}$, then 
$\sigma''$ is a model for $\driple{\liftphi_3}{\statephi_3}$.
\begin{align*}
\forall \sigma,\sigma'.~(\sigma,\sigma') \in \llbracket \statement_1;\statement_2 \rrbracket_{\contracts,\knowledge} \rightarrow
\exists \sigma''.~&(\sigma,\sigma'') \in \llbracket \statement_1 \rrbracket_{\contracts,\knowledge}  \wedge (\sigma'',\sigma') \in \llbracket \statement_2 \rrbracket_{\contracts,\knowledge} \wedge \\
& \left(\sigma \models_\knowledge \driple{\liftphi_1}{\statephi_1} \rightarrow \sigma'' \models_\knowledge \driple{\liftphi_3}{\statephi_3}\right)
\end{align*}
By applying the second premise we get that as $\sigma''$ is a model for $\sdriple{\liftphi_3}{\statephi_3}$, then 
$\sigma'$ is a model for $\driple{\liftphi_2}{\statephi_2}$.
\begin{align*}
\forall \sigma,\sigma'.~(\sigma,\sigma') \in \llbracket \statement_1;\statement_2 \rrbracket_{\contracts,\knowledge} \rightarrow
\exists \sigma''.~&(\sigma,\sigma'') \in \llbracket \statement_1 \rrbracket_{\contracts,\knowledge}  \wedge (\sigma'',\sigma') \in \llbracket \statement_2 \rrbracket_{\contracts,\knowledge} \wedge \\
& \left(\sigma'' \models_\knowledge \driple{\liftphi_3}{\statephi_3} \rightarrow \sigma' \models_\knowledge \driple{\liftphi_2}{\statephi_2} \right)
\end{align*}
As the existence of $\sigma''$ is ensured by the definition of composition, simplification of the formula gives us the validity of the conclusion
\begin{align*}
    \forall (\sigma,\sigma') \in \llbracket \statement_1;\statement_2 \rrbracket_{\contracts,\knowledge}.~\left(\sigma \models_\knowledge \driple{\liftphi_1}{\statephi_1} \rightarrow \sigma' \models_\knowledge \driple{\liftphi_2}{\statephi_2} \right)
\end{align*}
\item Rule \rulename{branch}.

We can assume that the following two conditions hold from the premises
\begin{align*}
    &\forall (\sigma, \sigma') \in \denotCK{\statement_1}.\, 
\left(\state \models_\knowledge \driple{\emptyset}{\statephi \wedge \mathtt{expr}}
\rightarrow \state' \models_\knowledge \driple{\liftphi}{\statephi}\right)\\
&\forall (\sigma, \sigma') \in \denotCK{\statement_2}.\, 
\left(\state \models_\knowledge \driple{\emptyset}{\statephi \wedge \neg\mathtt{expr}}
\rightarrow \state' \models_\knowledge \driple{\liftphi}{\statephi}\right)
\end{align*}
These we can combine into 
\begin{align*}
    \forall (\sigma, \sigma').\ &\sigma \models \mathsf{expr} \wedge (\sigma, \sigma') \in \denotCK{\statement_1}.\rightarrow
\left(\state \models_\knowledge \driple{\emptyset}{\statephi \wedge \mathtt{expr}}
\rightarrow \state' \models_\knowledge \driple{\liftphi}{\statephi}\right)\ \wedge\\
&\sigma \models \neg\mathsf{expr} \wedge (\sigma, \sigma') \in \denotCK{\statement_2} \rightarrow 
\left(\state \models_\knowledge \driple{\emptyset}{\statephi \wedge \neg\mathtt{expr}}
\rightarrow \state' \models_\knowledge \driple{\liftphi}{\statephi}\right)
\end{align*}
By definition of $\mathsf{cond}$ we have exactly validity of the branching statement.
\begin{align*}
    &\forall (\sigma, \sigma') \in \denotCK{\mathbf{if}(\mathtt{expr})\mathbf{then}~\statement_1~\mathbf{else}~\statement_2~\mathbf{fi}}.\, 
\left(\state \models_\knowledge \driple{\emptyset}{\statephi}
\rightarrow \state' \models_\knowledge \driple{\liftphi}{\statephi}\right)
\end{align*}
\item Rule \rulename{loop}. 
Follows directly from the definition of $\denotCK{\mathbf{while}~(\mathtt{e})~\mathbf{do}~\statement~\mathbf{od}}$. Note that we only consider terminating runs, so the definition is indeed well-founded. 
\item Rule \rulename{contract}. 
This is the usual contract rule and its soundness follows directly from the relativized program semantics.
\item Rule \rulename{var}.
\begin{align*}
    \forall (\sigma,\sigma')\in \llbracket \mathtt{v}~\mathtt{:=}~\mathtt{expr} \rrbracket_{\contracts,\knowledge}.&~\sigma \models_\knowledge\driple{\emptyset}{\statephi[\mathtt{v} \setminus \mathtt{expr}]} \rightarrow \sigma' \models_\knowledge \driple{\liftphi}{\statephi}\\
    \iff\forall (\sigma,\sigma')\in \llbracket \mathtt{v}~\mathtt{:=}~\mathtt{expr} \rrbracket_{\contracts,\knowledge}.&~\sigma,\mathcal{I} \models \statephi[\mathtt{v} \setminus \mathtt{expr}] \wedge \liftstate(\sigma),\liftspec(\statephi[\mathtt{v} \setminus \mathtt{expr}]) \models^\knowledge \emptyset\\ &\rightarrow \sigma',\mathcal{I} \models \statephi \wedge \liftstate(\sigma),\liftspec(\statephi) \models^\knowledge \liftphi\\
    \iff\forall (\sigma,\sigma')\in \llbracket \mathtt{v}~\mathtt{:=}~\mathtt{expr} \rrbracket_{\contracts,\knowledge}.&~\sigma,\mathcal{I} \models \statephi[\mathtt{v} \setminus \mathtt{expr}]\\ &\rightarrow \sigma',\mathcal{I} \models \statephi \wedge \liftstate(\sigma),\liftspec(\statephi) \models^\knowledge \liftphi\\
    \iff\forall \sigma.&~\sigma,\mathcal{I} \models \statephi[\mathtt{v} \setminus \mathtt{expr}]\\ &\rightarrow \sigma[\mathtt{v} \rightarrow \arexdenot{\mathtt{expr}}_\sigma],\mathcal{I} \models \statephi \wedge \liftstate(\sigma),\liftspec(\statephi) \models^\knowledge \liftphi\\
    \iff\forall \sigma.&~\sigma,\mathcal{I} \models \statephi[\mathtt{v} \setminus \mathtt{expr}] \rightarrow \sigma[\mathtt{v} \rightarrow \arexdenot{\mathtt{expr}}_\sigma],\mathcal{I} \models \statephi\tag{A}\\
    &\wedge \sigma,\mathcal{I} \models \statephi[\mathtt{v} \setminus \mathtt{expr}] \rightarrow\liftstate(\sigma),\liftspec(\statephi) \rightarrow \models^\knowledge \liftphi\tag{B}
\end{align*}
Condition (A) is the standard rule for assignment in the Hoare calculus, while condition (B) is implied by the premise.\qed
\end{itemize}

\end{proof}}

\begin{figure}[bt]
\begin{minipage}{.5\textwidth}
\begin{prooftree}
\AxiomC{$\liftspec(\statephi) \models^\knowledge \liftphi$}
\LeftLabel{\rulename{var}}
\UnaryInfC{$\contracts,\knowledge \vdash \driple{\emptyset}{\statephi[\mathtt{v} \setminus \mathtt{expr}]} \mathtt{v}~:=~\mathtt{expr} \driple{\liftphi}{\statephi}$}
\end{prooftree}
\end{minipage}
\begin{minipage}{.5\textwidth}
\begin{prooftree}
\AxiomC{$\ $}
\LeftLabel{\rulename{skip}}
\UnaryInfC{$\contracts,\knowledge \vdash \driple{\liftphi}{\statephi} \mathsf{skip} \driple{\liftphi}{\statephi}$}
\end{prooftree}
\end{minipage}

\begin{prooftree}
\AxiomC{$\contracts,\knowledge \vdash \driple{\emptyset}{\statephi \wedge \mathtt{expr}}\statement_1\driple{\liftphi}{\statephi}$}
\AxiomC{$\contracts,\knowledge \vdash \driple{\emptyset}{\statephi \wedge \neg\mathtt{expr}}\statement_2\driple{\liftphi}{\statephi}$}
\LeftLabel{\rulename{branch}}
\BinaryInfC{$\contracts,\knowledge \vdash \driple{\emptyset}{\statephi}\mathbf{if}~(\mathtt{expr})~\mathbf{then}~\statement_1~\mathbf{else}~\statement_2~\mathbf{fi}~\driple{\liftphi}{\statephi}$}
\end{prooftree}

\begin{prooftree}
\AxiomC{$\contracts,\knowledge \vdash \driple{\liftphi_1}{\statephi_1}\mathbf{if}~(\mathtt{expr})~\mathbf{do}~\statement;~\mathbf{while}~(\mathtt{expr})~\mathbf{do}~\statement~\mathbf{od}~\mathbf{else}~\mathbf{skip}~\mathbf{fi}\driple{\liftphi_2}{\statephi_2}$}
\LeftLabel{\rulename{loop}}
\UnaryInfC{$\contracts,\knowledge \vdash \driple{\liftphi_1}{\statephi_1}\mathbf{while}~(\mathtt{expr})~\mathbf{do}~\statement~\mathbf{od}~\driple{\liftphi_2}{\statephi_2}$}
\end{prooftree}

\begin{prooftree}
\AxiomC{$\ $}
\LeftLabel{\rulename{contract}}
\UnaryInfC{$\contracts,\knowledge \vdash \mathsf{Pre}(\contracts, \mathtt{p}, \mathtt{expr})~\mathtt{p}(\mathtt{e})~\mathsf{Post}(\contracts, \mathtt{p}, \mathtt{expr})$}
\end{prooftree}

\begin{prooftree}
\AxiomC{$\contracts,\knowledge \vdash \driple{\liftphi_1}{\statephi_1} \statement_1 \driple{\liftphi_3}{\statephi_3}$}
\AxiomC{$\contracts,\knowledge \vdash \driple{\liftphi_3}{\statephi_3} \statement_2 \driple{\liftphi_2}{\statephi_2}$}
\LeftLabel{\rulename{seq}}
\BinaryInfC{$\contracts,\knowledge \vdash \driple{\liftphi_1}{\statephi_1} \statement_1;~\statement_2 \driple{\liftphi_2}{\statephi_2}$}
\end{prooftree}
\vspace{-4mm}
    \caption{Rules for weakest precondition reasoning with lifted assertions.}
    \label{fig:calc1}
\end{figure}

Given the above rules, we can easily combine several operations to derive sound rules that operate in the domain as well.  One simple way is to merely lift before and after the statement, such as in the following derived rule.

\begin{prooftree}
\AxiomC{$\ $}
\LeftLabel{\rulename{lift-var}}
\UnaryInfC{$\contracts,\knowledge \vdash \driple{\liftspec(\statephi[\mathtt{v} \setminus \mathtt{expr}])}{\statephi[\mathtt{v} \setminus \mathtt{expr}]} \mathtt{v}~\mathtt{:=}~\mathtt{expr} \driple{\liftspec(\statephi)}{\statephi}$}
\end{prooftree}

While \rulename{lift-var} is sound, it does not transfer any information from the domain postcondition; the domain precondition is computed by a function of only the state precondition -- it is, thus, not computing the weakest domain-precondition. 

\begin{proposition}\label{prop1}
Rule \rulename{lift-var} is sound.
\end{proposition}
\techrep{
\begin{proof}
We derive
\begin{prooftree}
\AxiomC{$\ $}
\UnaryInfC{$\driple{\liftspec(\statephi[\mathtt{v} \setminus \mathtt{e}])}{\statephi[\mathtt{v} \setminus \mathtt{e}]} \rightarrow_\knowledge \driple{\emptyset}{\statephi[\mathtt{v} \setminus \mathtt{e}]}$}
\AxiomC{$\ $}
\UnaryInfC{$\liftspec(\statephi) \models_\knowledge \liftspec(\statephi)$}
\RightLabel{\rulename{var}}
\UnaryInfC{$\contracts,\knowledge \vdash \driple{\emptyset}{\statephi[\mathtt{v} \setminus \mathtt{e}]} \mathtt{v}~\mathtt{:=}~\mathtt{e} \driple{\liftspec(\statephi)}{\statephi}$}
\RightLabel{\rulename{cons}}
\BinaryInfC{$\contracts,\knowledge \vdash \driple{\liftspec(\statephi[\mathtt{v} \setminus \mathtt{e}])}{\statephi[\mathtt{v} \setminus \mathtt{e}]} \mathtt{v}~\mathtt{:=}~\mathtt{e} \driple{\liftspec(\statephi)}{\statephi}$}
\end{prooftree}
Consequently, by Thm.~\ref{thm:sound}, rule \rulename{lift-var} is sound.\qed
\end{proof}}

Using the mechanisms around the lifted core, we can give more precise versions of the rules for statements. 
Let $\mathsf{DPre}(\liftphi,\statephi)$ be the domain knowledge constructed by abducting a lifted core from the domain postcondition ($\abduct(\liftphi)$), delifting it into the state logic (via $\liftspec^{-1}$), performing the substitution on the delifted core and the computation specification ($[\mathtt{v} \setminus e]$),
and lifting the result back into the state logic ($\liftspec$), thus realizing one full cycle of the information flow in Fig.~\ref{fig:relations}:
\[ \mathsf{DPre}(\liftphi,\statephi) = \liftspec\Big(\big(\statephi\wedge\liftspec^{-1}(\abduct_\knowledge(\liftphi))\big)[\mathtt{v} \setminus \mathtt{expr}]\Big)\]

However, for soundness it remains to show that the generated core is indeed implied by the domain specification.
As discussed, this may not be the case if we use abduction for core generation.
In this case, the open proof branches witness the abducted core and can be examined by the user. The following rule, for example, does so by integrating all steps to derive the domain precondition too.
\begin{prooftree}
\AxiomC{$\liftphi \models^\knowledge \abduct_\knowledge(\liftphi)$}
\AxiomC{$\liftspec(\statephi)\models^\knowledge \liftphi$}
\LeftLabel{\rulename{total}}
\BinaryInfC{$\contracts,\knowledge \vdash \driple{\mathsf{DPre}(\liftphi,\statephi)}{\statephi\wedge\liftspec^{-1}(\abduct_\knowledge(\liftphi))} \mathtt{v}~\mathtt{:=}~\mathtt{expr} \driple{\liftphi}{\statephi}$}
\end{prooftree}
\begin{proposition}\label{prop2}
    Rule \rulename{total} is sound.
\end{proposition}
\techrep{\begin{proof}
Let $\widehat{\liftphi,\statephi} = \statephi\wedge\liftspec^{-1}(\abduct(\liftphi))$.
We derive

\begin{prooftree}
\AxiomC{$\ast$}
\AxiomC{$\liftspec(\statephi) \models^\knowledge \liftphi$}
\UnaryInfC{$\liftspec(\widehat{\liftphi,\statephi}) \models^\knowledge \liftphi$}
\RightLabel{\rulename{var}}
\UnaryInfC{$\contracts,\knowledge \vdash \driple{\emptyset}{\widehat{\liftphi,\statephi}[\mathtt{v} \setminus e]} \mathtt{v}~\mathtt{:=}~\mathtt{e} \driple{\liftphi}{\widehat{\liftphi,\statephi}}$}
\AxiomC{$\liftphi \models^\knowledge \abduct(\liftphi)$}
\UnaryInfC{$\driple{\liftphi}{\widehat{\liftphi,\statephi}} \rightarrow_\knowledge \driple{\liftphi,\abduct(\liftphi)}{\widehat{\liftphi,\statephi}}$}
\LeftLabel{\rulename{cons}}
\BinaryInfC{$\contracts,\knowledge \vdash \driple{\emptyset}{\widehat{\liftphi,\statephi}[\mathtt{v} \setminus e]} \mathtt{v}~\mathtt{:=}~\mathtt{e} \driple{\liftphi,\abduct(\liftphi)}{\widehat{\liftphi,\statephi}}$}
\RightLabel{\rulename{post-inv}}
\UnaryInfC{$\contracts,\knowledge \vdash \driple{\emptyset}{\widehat{\liftphi,\statephi}[\mathtt{v} \setminus e]} \mathtt{v}~\mathtt{:=}~\mathtt{e} \driple{\liftphi,\abduct(\liftphi)}{\statephi}$}
\RightLabel{\rulename{post-abs}}
\UnaryInfC{$\contracts,\knowledge \vdash \driple{\emptyset}{\widehat{\liftphi,\statephi}[\mathtt{v} \setminus e]} \mathtt{v}~\mathtt{:=}~\mathtt{e} \driple{\liftphi}{\statephi}$}
\RightLabel{\rulename{cons}}
\BinaryInfC{$\contracts,\knowledge \vdash \driple{\liftspec\Big(\widehat{\liftphi,\statephi}[\mathtt{v} \setminus e]\Big)}{\widehat{\liftphi,\statephi}[\mathtt{v} \setminus e]} \mathtt{v}~\mathtt{:=}~\mathtt{e} \driple{\liftphi}{\statephi}$}
\end{prooftree}

\begin{prooftree}
\AxiomC{$\ $}
\UnaryInfC{$\driple{\liftspec\Big(\widehat{\liftphi,\statephi}[\mathtt{v} \setminus e]\Big)}{\widehat{\liftphi,\statephi}[\mathtt{v} \setminus e]} \rightarrow_\knowledge \driple{\emptyset}{\widehat{\liftphi,\statephi}[\mathtt{v} \setminus e]}$}
\UnaryInfC{$\ast$}
\end{prooftree}

Consequently, by Thm.~\ref{thm:sound}, rule \rulename{total-var} is sound.\qed
\end{proof}}
\paragraph{Example.}
\begin{figure}[b!t]
\begin{align*}
    \knowledge = \big\{&
\mathtt{SmallCar} \equiv \mathtt{HasTwoDoors} \sqcap \mathtt{HasFourWheels},\quad
\mathtt{HasBody} \sqsubseteq \mathtt{HasChassis} \sqsubseteq \mathtt{Car}\\
&\exists \mathtt{doors}. \exists \mathtt{hasValue}. 2 \ensuremath{\equiv} \mathtt{HasTwoDoors},\quad
\exists \mathtt{wheels}. \exists \mathtt{hasValue}. 4 \ensuremath{\equiv} \mathtt{HasFourWheels}\\
&\exists \mathtt{body}. \mathtt{NonZero} \ensuremath{\equiv}\mathtt{HasBody}\quad
\neg\exists \mathtt{hasValue}.0 \equiv \mathtt{NonZero},\quad
\mathtt{HasChassis}(c),\\
&\mathtt{wheels}(c, \mathtt{wheelsVar}) \mathtt{body}(c, \mathtt{bodyVar}),\quad\mathtt{doors}(c, \mathtt{doorsField})
\big\}\\
    &\mathtt{Pre}_\contracts(\mathtt{addWheels},\mathtt{nrWheels}) = \sdriple{-}{\mathtt{nrWheel} \doteq 4}\\
    &\mathtt{Post}_\contracts(\mathtt{addWheels},\mathtt{nrWheels}) =\sdriple{\mathtt{HasFourWheels}(c)}{-} \\
    &\mathtt{Pre}_\contracts(\mathtt{assembly}) = \sdriple{-}{\mathtt{nrDoors} \doteq 4 \wedge \mathtt{id} \neq 0}\\
    &\mathtt{Post}_\contracts(\mathtt{assembly}) = \sdriple{\mathtt{SmallCar}(c)}{-}\\[-10mm]
\end{align*}
    \caption{Domain knowledge and specification of the running example.}
    \label{fig:ex:know}
\end{figure}

Let us now return to the assembly line, where we can now finally give a formal proof of our running example.  The domain knowledge $\knowledge$ are the axioms from Sec.~\ref{sec:example}, given in Fig.~\ref{fig:ex:know} together with the contracts for all procedures.
The proof for the contract of $\mathtt{addWheels}$ is given below. 
\[
\liftspec(\mathtt{wheel} \doteq n) = \{\mathtt{HasValue}(\mathtt{wheelsVar},n)\} 
\]
Rule \rulename{post-abd} is applied first and adds $\mathtt{HasValue}(\mathtt{wheelsVar},4)$ in the kernel generation through deduction -- this follows from the \emph{equivalence} axiom for $\mathtt{HasFourWheels}$, as well as $\mathtt{wheels}(c, \mathtt{wheelsVar})$. The second applied rule is \rulename{post-inv}, where this axiom is used to recover $\mathtt{wheel} \doteq 4$.
The third applied rule is \rulename{new-var}, where we must show that the post-condition is strongly consistent.

\begin{prooftree}
\AxiomC{}
\UnaryInfC{$\mathtt{hasValue}(\mathtt{wheelsVar},4) \models^\knowledge \mathtt{HasFourWheels}(c), \mathtt{hasValue}(\mathtt{wheelsVar},4)$}
\UnaryInfC{$\contracts, \knowledge \vdash 
\driple{-}{\mathtt{nrWheels} \doteq 4} 
\mathtt{wheels}  := \mathtt{nrWheels}
\driple{\mathtt{HasFourWheels}(c), \mathtt{hasValue}(\mathtt{wheelsVar},4)}{\mathtt{wheels} \doteq 4}$}
\UnaryInfC{$\contracts, \knowledge \vdash 
\driple{-}{\mathtt{nrWheels} \doteq 4} 
\mathtt{wheels}  := \mathtt{nrWheels}
\driple{\mathtt{HasFourWheels}(c), \mathtt{hasValue}(\mathtt{wheelsVar},4)}{-}$}
\UnaryInfC{$\contracts, \knowledge \vdash 
\driple{-}{\mathtt{nrWheels} \doteq 4} 
\mathtt{wheels}  := \mathtt{nrWheels}
\driple{\mathtt{HasFourWheels}(c)}{-}$}
\end{prooftree}

\conference{
For brevity, the proof for the contract of $\mathtt{assembly}$ is given in the appendix. 
That proof also includes a stronger contract for \texttt{addWheels} that is required for framing, but bears no insights into the interplay of program and domain.
}
\techrep{

In the following we give a full proof of the running example of this work. To keep the deduction trees readable, we use the following abbreviations.
Let $sw$ be $\mathtt{wheels}  := \mathtt{nrWheels}$ and $\Delta^1$ be $\mathtt{HasFourWheels}(c), \mathtt{hasValue}(\mathtt{wheelsVar},4)$. 
Let $sd$ be $\mathtt{doors := nrDoors}$ and $\Delta^2$ be 
\[\mathtt{hasValue}(\mathtt{wheelsVar},4),\mathtt{hasValue}(\mathtt{doorsVar},4),\mathtt{nonZero}(\mathtt{idVar})\]

We prove the contract of \texttt{addWheels} again, but now with a framing condition $\mathtt{nrDoors} \doteq 4 \wedge \mathtt{id} \neq 0$ that does not influence the form of the proof.

\noindent
\scalebox{1}{
\parbox{1cm}{
\begin{prooftree}
\AxiomC{\ }
\LeftLabel{DL}
\UnaryInfC{$\mathtt{hasValue}(\mathtt{wheelsVar},4) \models^\knowledge \Delta^1$}
\LeftLabel{\rulename{var}}
\UnaryInfC{$\contracts, \knowledge \vdash 
\driple{-}{\mathtt{nrWheels} \doteq 4 \wedge \mathtt{nrDoors} \doteq 4 \wedge \mathtt{id} \neq 0} 
sw
\driple{\Delta^1}{\mathtt{wheels} \doteq 4 \wedge \mathtt{nrDoors} \doteq 4 \wedge \mathtt{id} \neq 0}$}
\LeftLabel{\rulename{post-inv}}
\UnaryInfC{$\contracts, \knowledge \vdash 
\driple{-}{\mathtt{nrWheels} \doteq 4 \wedge \mathtt{nrDoors} \doteq 4 \wedge \mathtt{id} \neq 0} 
sw
\driple{\Delta^1}{\mathtt{nrDoors} \doteq 4 \wedge \mathtt{id} \neq 0}$}
\AxiomC{(1)}
\LeftLabel{\rulename{post-core}}
\BinaryInfC{$\contracts, \knowledge \vdash 
\driple{-}{\mathtt{nrWheels} \doteq 4 \wedge \mathtt{nrDoors} \doteq 4 \wedge \mathtt{id} \neq 0} 
sw
\driple{\mathtt{HasFourWheels}(c)}{ \mathtt{nrDoors} \doteq 4 \wedge \mathtt{id} \neq 0}$}
\end{prooftree}
}}

\noindent where the side branch~(1) is closed as follows.
\medskip
\begin{prooftree}
\AxiomC{\ }
\RightLabel{DL}
\UnaryInfC{$\mathtt{HasFourWheels}(c) \models_\knowledge \mathtt{hasValue}(\mathtt{wheelsVar},4)$}
\UnaryInfC{(1)}
\end{prooftree}

Next, we turn to \texttt{assembly}. First, we split the three statements with fitting intermediate assertions.
These assertions are chosen according to the contract of the \texttt{addWheels} call in the middle statement.
The first statement is trivial, as it involves no domain specification at all.

\noindent
\scalebox{.925}{
\parbox{1cm}{
\begin{prooftree}
\AxiomC{\ }
\LeftLabel{DL}
\UnaryInfC{$\emptyset \models^\knowledge \emptyset$}
\LeftLabel{\rulename{var}}
\UnaryInfC{$\contracts, \knowledge \vdash 
\driple{-}{\mathtt{nrDoors} \doteq 4 \wedge \mathtt{id} \neq 0} 
  \mathtt{bodyId := id;}
\driple{-}{\mathtt{nrDoors} \doteq 4 \wedge \mathtt{bodyId} \neq 0}$}
\AxiomC{(2)}
\RightLabel{\rulename{seq}}
\BinaryInfC{$\contracts, \knowledge \vdash 
\driple{-}{\mathtt{nrDoors} \doteq 4 \wedge \mathtt{id} \neq 0} 
  \mathtt{bodyId := id; addWheels(4);}
\driple{\mathtt{HasFourWheels}(c)}{\mathtt{nrDoors} \doteq 4 \wedge \mathtt{id} \neq 0}$}
\AxiomC{(3)}
\BinaryInfC{$\contracts, \knowledge \vdash 
\driple{-}{\mathtt{nrDoors} \doteq 4 \wedge \mathtt{id} \neq 0} 
  \mathtt{bodyId := id; addWheels(4); doors := nrDoors;}
\driple{\mathtt{SmallCar}(c)}{-}$}
\end{prooftree}
}}

Second, we prove the correctness of the procedure call. This is a simple manner of fitting it into the 
required syntactic form of the specification using the consequence rule.

\noindent
\scalebox{.825}{
\parbox{1cm}{
\begin{prooftree}
\AxiomC{(2')}
\AxiomC{\ }
\LeftLabel{\rulename{contract}}
\UnaryInfC{$\contracts, \knowledge \vdash 
\driple{-}{\mathtt{nrDoors} \doteq 4 \wedge \mathtt{bodyId} \neq 0 \wedge 4 \doteq 4}
  \mathtt{addWheels(4);}
\driple{\mathtt{HasFourWheels}(c)}{\mathtt{nrDoors} \doteq 4 \wedge \mathtt{id} \neq 0}$}
\AxiomC{(2'')}
\RightLabel{\rulename{cons}}
\TrinaryInfC{$\contracts, \knowledge \vdash 
\driple{-}{\mathtt{nrDoors} \doteq 4 \wedge \mathtt{bodyId} \neq 0}
  \mathtt{addWheels(4);}
\driple{\mathtt{HasFourWheels}(c)}{\mathtt{nrDoors} \doteq 4 \wedge \mathtt{id} \neq 0}$}
\UnaryInfC{(2)}
\end{prooftree}
}}

\noindent where the side branches~(2') and~(2'') are closed as follows.
\begin{prooftree}
\AxiomC{\ }
\LeftLabel{}
\UnaryInfC{$
\driple{-}{\mathtt{nrDoors} \doteq 4 \wedge \mathtt{bodyId} \neq 0 \wedge 4 \doteq 4} 
\rightarrow_\knowledge
\driple{-}{\mathtt{nrDoors} \doteq 4 \wedge \mathtt{bodyId} \neq 0}$}
\UnaryInfC{(2')}
\end{prooftree}

\begin{prooftree}
\AxiomC{\ }
\LeftLabel{}
\UnaryInfC{$
\driple{\mathtt{HasFourWheels}(c)}{\mathtt{nrDoors} \doteq 4 \wedge \mathtt{id} \neq 0}
\rightarrow_\knowledge
\driple{\mathtt{HasFourWheels}(c)}{\mathtt{nrDoors} \doteq 4 \wedge \mathtt{id} \neq 0}$}
\UnaryInfC{(2'')}
\end{prooftree}

Third, we turn to the final statement. This is the most interesting step, as it has domain specification in both pre- and post-condition. As the \rulename{var} rule removes the domain precondition, we need 
to generate the core for both pre- and post-condition, and then recover the state specification from it.

\noindent
\scalebox{.85}{
\parbox{1cm}{
\begin{prooftree}
\AxiomC{(4)}
\RightLabel{\rulename{post-inv}}
\UnaryInfC{$\contracts, \knowledge \vdash 
\driple{\mathtt{HasFourWheels}(c),\mathtt{hasValue}(\mathtt{wheelsVar},4)}{\mathtt{nrDoors} \doteq 4 \wedge \mathtt{id} \neq 0 \wedge \mathtt{wheel} \doteq 4}
sd
\driple{\mathtt{SmallCar}(c),\Delta^2}{-}$}
\RightLabel{\rulename{pre-inv}}
\UnaryInfC{$\contracts, \knowledge \vdash 
\driple{\mathtt{HasFourWheels}(c),\mathtt{hasValue}(\mathtt{wheelsVar},4)}{\mathtt{nrDoors} \doteq 4 \wedge \mathtt{id} \neq 0}
sd
\driple{\mathtt{SmallCar}(c),\Delta^2}{-}$}
\AxiomC{(3'')}
\LeftLabel{\rulename{pre-core}}
\BinaryInfC{$\contracts, \knowledge \vdash 
\driple{\mathtt{HasFourWheels}(c)}{\mathtt{nrDoors} \doteq 4 \wedge \mathtt{id} \neq 0}
sd
\driple{\mathtt{SmallCar}(c),\Delta^2}{-}$}
\AxiomC{(3')}
\LeftLabel{\rulename{post-core}}
\BinaryInfC{$\contracts, \knowledge \vdash 
\driple{\mathtt{HasFourWheels}(c)}{\mathtt{nrDoors} \doteq 4 \wedge \mathtt{id} \neq 0}
sd
\driple{\mathtt{SmallCar}(c)}{-}$}
\end{prooftree}
}}

Using the consequence rule we then remove the domain precondition, and again apply the \rulename{var} rule.
\begin{prooftree}
\AxiomC{\ }
\RightLabel{DL}
\UnaryInfC{$\mathtt{nrDoors} \doteq 4 \wedge \mathtt{id} \neq 0 \wedge \mathtt{wheel} \doteq 4 \models_\knowledge \mathtt{SmallCar}(c),\Delta^2$}
\RightLabel{\rulename{var}}
\UnaryInfC{$\contracts, \knowledge \vdash 
\driple{-}{\mathtt{nrDoors} \doteq 4 \wedge \mathtt{id} \neq 0 \wedge \mathtt{wheel} \doteq 4}
sd
\driple{\mathtt{SmallCar}(c),\Delta^2}{\mathtt{nrDoors} \doteq 4 \wedge \mathtt{id} \neq 0 \wedge \mathtt{wheel} \doteq 4}$}
\AxiomC{(4')}
\RightLabel{\rulename{cons}}
\BinaryInfC{$\contracts, \knowledge \vdash 
\driple{\mathtt{HasFourWheels}(c),\mathtt{hasValue}(\mathtt{wheelsVar},4)}{\mathtt{nrDoors} \doteq 4 \wedge \mathtt{id} \neq 0 \wedge \mathtt{wheel} \doteq 4}
sd
\driple{\mathtt{SmallCar}(c),\Delta^2}{\mathtt{doors} \doteq 4 \wedge \mathtt{id} \neq 0 \wedge \mathtt{wheel} \doteq 4}$}
\UnaryInfC{(4)}
\end{prooftree}

\noindent The final side branches are closed as follows.

\begin{prooftree}
\AxiomC{\ }
\LeftLabel{DL}
\UnaryInfC{$\mathtt{SmallCar}(c) \models^\knowledge \Delta^2$}
\UnaryInfC{(3')}
\end{prooftree}
\begin{prooftree}
\AxiomC{\ }
\LeftLabel{DL}
\UnaryInfC{$\mathtt{HasFourWheels}(c) \models^\knowledge\mathtt{hasValue}(\mathtt{wheelsVar},4)$}
\UnaryInfC{(3'')}
\end{prooftree}

\begin{prooftree}
\AxiomC{\ }
\LeftLabel{}
\UnaryInfC{$\driple{\mathtt{HasFourWheels}(c),\mathtt{hasValue}(\mathtt{wheelsVar},4)}{\mathtt{nrDoors} \doteq 4 \wedge \mathtt{id} \neq 0 \wedge \mathtt{wheel} \doteq 4} \rightarrow\knowledge \driple{-}{\mathtt{nrDoors} \doteq 4 \wedge \mathtt{id} \neq 0 \wedge \mathtt{wheel} \doteq 4}$}
\UnaryInfC{(4')}
\end{prooftree}
}

\section{Related Work}\label{sec:related}

While specification is a long-standing challenge for deductive verification~\cite{DBLP:journals/corr/abs-1211-6186,DBLP:conf/vstte/Rozier16,DBLP:series/lncs/HahnleH19}, integration of description logics, or related technologies, such as the semantic web stack, into deductive verification of mainstream programming languages has not been explored. However, the integration of description logics directly into programming languages, for example through language-integrated queries or epistemic operators, has been investigated, as have been model checking approaches for such integrations.

(Con)Golog~\cite{DBLP:journals/jlp/LevesqueRLLS97,DBLP:journals/ai/GiacomoLL00} is an action programming language based on the situation calculus, designed to program agents that must access the current situation of their dynamic context.
Golog has been connected with description logics to achieve decidable verification~\cite{ZarriessC15,DBLP:phd/dnb/Zarriess18} of temporal logic properties, based on abstraction into a system where model checking is decidable~\cite{DBLP:conf/frocos/BaaderZ13}. To do so, the external world is modeled as a description logic model. In contrast to the work on Golog verification, this work targets mainstream imperative programming, where description logics are used only for specification, and focuses on the interplay of specification and deductive calculus. As we aim for better specification of the domain constraints, questions of decidability are of lesser interest here.

Knowledge and action bases~\cite{DBLP:journals/jair/HaririCMGMF13} allow programs to access and manipulate a knowledge base using two abstraction operators, \texttt{ask} and \texttt{tell}. Again, verification of temporal properties based on model checking has been considered~\cite{DBLP:journals/jair/HaririCMGMF13} with a focus on decidability~\cite{DBLP:conf/aaai/CalvaneseGMM23,DBLP:conf/jelia/CalvaneseCMS14}, but no deductive system. Similarly, knowledge-based programs~\cite{DBLP:books/mit/FHMV1995} are based on epistemic operators and have only been considered for analysis of simple temporal properties~\cite{DBLP:conf/esop/KnappMR23}.

The original work on semantically lifted programs~\cite{DBLP:conf/esws/KamburjanKSJG21} uses integrated queries to access the lifted state. A similar mechanism is used by the probabilistic, ontologized programs of Dubslaff et al.~\cite{DBLP:journals/fac/DubslaffKT21}, which also give a model checker for temporal properties based on SPIN. For semantically lifted programs, a type system is given in~\cite{DBLP:conf/dlog/KamburjanK21}, which is using description logic entailments to verify graph query containments that ensure safety of the language-integrated queries.
Leinberger et al.~\cite{DBLP:conf/semweb/LeinbergerSSLS19} also give a type system, but base their system not on  liftings and graph queries, but on a tight integration of the class systems and graph shapes~\cite{DBLP:conf/esop/LeinbergerLS17}, which are again reduced to description logic entailments~\cite{DBLP:conf/semweb/LeinbergerSRLS20}.


\section{Conclusion}\label{sec:conclude}
This work presents a two-tier Hoare logic that integrates description logic specification over a semantically lifted program. At its heart, it introduces semantical lifting of specifications, which must be compatible with the lifting of states. In the calculus, kernel-generation generates axioms in a certain signature to produce information about the program state from the domain view. We aim to continue this work to leverage the pragmatics of knowledge representation with description logic to program specification and their deductive verification.

The conditions on the lifting and its integration into the calculus are the main result, and the used programming language is consequently kept minimal. Thus, questions of expressive power and complexity are left for future work. 

\clearpage
\bibliographystyle{splncs04}
\bibliography{ref}

  \conference{\FloatBarrier
  \clearpage
  \appendix
  Due to space constraints, some formal definitions, the proofs and the full example are given in this appendix. 
In case of acceptance, this appendix will be published as a technical report on arXiv and referenced from the paper.
\section{Full Definitions}

\subsection*{State Logic (Def.~\ref{def:sl})}

Let $\mathsf{V}$ be the set of program variables. 
A \emph{program state} $\state: \mathsf{V} \rightarrow \mathsf{Val}$ is a mapping from variables to values.
Let $\mathtt{S}$ denote the set of all program states.
Let $\Sigma= \langle \mathtt{V}, \mathtt{F}, \mathtt{F}_d, \mathtt{P}\rangle$ be a state signature.
State formulas $\statephi$ are defined by the following grammar, where $v$ ranges over $\mathtt{V}$, $p$ over $\mathtt{P}$, and $f$ over $\mathtt{F} \cup \mathtt{F}_d$. The set of all state formulas over $\Sigma$ is denoted $\Phi(\Sigma)$.
\[
\statephi ::= \statephi \wedge \statephi ~|~ \neg \statephi ~|~t \doteq t~|~ p(\overline{t}) 
\qquad\qquad t ::= v ~|~ x ~|~ f(\overline{t})
\]

The semantics of the state logic $\sigma, \mathcal{I} \models \Phi$ is defined relative to a program state and a state interpretation, and is given in Fig.~\ref{fig:fol:sem}.
\begin{figure}[b]
\begin{align*}
    \sigma, I &\models \statephi_1 \wedge \statephi_2 \iff \sigma, I \models \statephi_1 \text{ and } \sigma, I \models \statephi_2 \\
    \sigma, I &\models \neg\statephi \iff \sigma, I \not\models \statephi\\
    \sigma, I &\models t_! \doteq t_2 \iff \mathit{val}_{\sigma, I }(t_1) = \mathit{val}_{\sigma, I }(t_2)\\
    \sigma, I &\models p(t_1,\dots,t_n) \iff I(p)(\mathit{val}_{\sigma, I }(t_1),\dots, \mathit{val}_{\sigma, I }(t_n))\\ \\
    \mathit{val}_{\sigma, I }(v) &= \sigma(v) \qquad 
    \mathit{val}_{\sigma, I }(f(t_1,\dots,t_n)) = I(f)(\mathit{val}_{\sigma, I }(t_1),\dots, \mathit{val}_{\sigma, I }(t_n))
\end{align*}
     \caption{Semantics of the state logic.}
     \label{fig:fol:sem}
 \end{figure}

\subsection*{Description Logic (Def.~\ref{def:dl})}

Let $\Sigma_d = \langle \mathtt{N}, \mathtt{R}, \mathtt{T}, \mathtt{A}\rangle$ be a domain signature.
The syntax of domain formulas $\delta$ is defined by the following grammar, where $A$ ranges over $\mathtt{A}$,   $R$ over $\mathtt{R}$, $T$ over $\mathtt{T}$,  $o$ over $\mathtt{N}$, and $n$ over literals from $\mathbb{Z}$. The set of all domain formulas over $\Sigma_d$ is denoted $\Delta(\Sigma_d)$. We use $\liftphi$ to range over sets of domain formulas.
\begin{align*}
    \delta ::=\>& C \sqsubseteq C ~|~ C(o) ~|~ R(o,o) ~|~ R(o,n)\\
    C ::=\>& \top ~|~ \bot ~|~ A ~|~ \neg C ~|~ C \sqcup C ~|~ C \sqcap C ~|~ \exists R.\, C ~|~ \forall R.\, C ~|~ \exists T.\, n ~|~ \forall T.\, n
\end{align*}
The semantics $\mathcal{I}_d \models \delta$ is defined relative to a domain interpretation, and is given in Fig.~\ref{fig:dl:sem}.
\begin{figure}[t]
\begin{align*}
    \mathit{val}_\mathcal{I}(\top) &= \nabla \qquad\quad~  \mathit{val}_\mathcal{I}(\bot) = \emptyset \\
    \mathit{val}_\mathcal{I}(A) &= \mathcal{I}(A) \qquad \mathit{val}_\mathcal{I}(\neg C) = \nabla \setminus \mathit{val}_\mathcal{I}(C) \\
    \mathit{val}_\mathcal{I}(C \sqcup D) &= \mathit{val}_\mathcal{I}(C) \cup \mathit{val}_\mathcal{I}(D) \qquad
    \mathit{val}_\mathcal{I}(C \sqcap D) = \mathit{val}_\mathcal{I}(C) \cap \mathit{val}_\mathcal{I}(D) \\
    \mathit{val}_\mathcal{I}(\exists R.\, C) &= \{x \in \nabla ~|~ \exists y.~(x,y) \in \mathcal{I}(R) \wedge y \in \mathcal{I}(C)\} \\
    \mathit{val}_\mathcal{I}(\forall R.\, C) &= \{x \in \nabla ~|~ \forall y.~(x,y) \in \mathcal{I}(R) \wedge y \in \mathcal{I}(C)\}\\
    \mathit{val}_\mathcal{I}(\exists T.\, d) &= \{x \in \mathbb{Z} ~|~ \exists y.~(x,y) \in \mathcal{I}(T) \wedge y \in \mathcal{I}(T)\} \\
    \mathit{val}_\mathcal{I}(\forall T.\, d) &= \{x \in \mathbb{Z} ~|~ \forall y.~(x,y) \in \mathcal{I}(T) \wedge y \in \mathcal{I}(T)\}\\ 
    \mathcal{I} \models C \subseteq D &\iff \mathit{val}_\mathcal{I}(C) \subseteq \mathit{val}_\mathcal{I}(D)\qquad
    \mathcal{I} \models C(o) \iff \mathcal{I}(o) \in \mathit{val}_\mathcal{I}(C)\\
    \mathcal{I} \models R(o,o') &\iff (\mathcal{I}(c),\mathcal{I}(o') \in \mathcal{I}(R)\qquad
    \mathcal{I} \models T(o,d) \iff (\mathcal{I}(c),\mathcal{I}(d)) \in \mathcal{I}(T)
\end{align*}
    \caption{Semantics of the domain logic.}
    \label{fig:dl:sem}
\end{figure}

\section{Proofs}
\subsection*{Lemma~\ref{lem:conditions}}

\begin{proof}
 Property $1.$ holds because $\alpha(\Delta)$ is a weakening and does not impose additional conditions. From Def.~\ref{def:abduct} we have
    \[\alpha_\knowledge(\Delta)\models_\knowledge \Delta \iff \forall \mathcal{I}.~\mathcal{I} \models_\knowledge\alpha_\knowledge(\Delta)\rightarrow \mathcal{I}\models_\knowledge \Delta\]
    So, the following relation holds by the semantics of the domain logic. \[\mathcal{I} \models_\knowledge\alpha_\knowledge(\Delta), \Delta~\rightarrow~ 
    \mathcal{I} \models_\knowledge \Delta\]
    This is trivially applied to lifted assertions, where $\mathcal{I}$ can be constructed from~$\sigma$.
    \[
    \sigma \models_\knowledge \driple{\liftphi,\alpha_\knowledge(\Delta)}{\Phi} \rightarrow 
    \sigma \models_\knowledge \driple{\liftphi}{\Phi} 
    \]
Properties $2.$ and $3.$ follow directly from Def.~\ref{def:compatible} and Def.~\ref{def:liftedasrt}.
\end{proof}

\subsection*{Theorem~\ref{thm:sound}}

\subsection*{Proposition~\ref{prop1}}

\begin{proof}
We derive
\begin{prooftree}
\AxiomC{$\ $}
\UnaryInfC{$\driple{\liftspec(\statephi[\mathtt{v} \setminus \mathtt{e}])}{\statephi[\mathtt{v} \setminus \mathtt{e}]} \rightarrow_\knowledge \driple{\emptyset}{\statephi[\mathtt{v} \setminus \mathtt{e}]}$}
\AxiomC{$\ $}
\UnaryInfC{$\liftspec(\statephi) \models_\knowledge \liftspec(\statephi)$}
\RightLabel{\rulename{var}}
\UnaryInfC{$\contracts,\knowledge \vdash \driple{\emptyset}{\statephi[\mathtt{v} \setminus \mathtt{e}]} \mathtt{v}~\mathtt{:=}~\mathtt{e} \driple{\liftspec(\statephi)}{\statephi}$}
\RightLabel{\rulename{cons}}
\BinaryInfC{$\contracts,\knowledge \vdash \driple{\liftspec(\statephi[\mathtt{v} \setminus \mathtt{e}])}{\statephi[\mathtt{v} \setminus \mathtt{e}]} \mathtt{v}~\mathtt{:=}~\mathtt{e} \driple{\liftspec(\statephi)}{\statephi}$}
\end{prooftree}
Consequently, by Thm.~\ref{thm:sound}, rule \rulename{lift-var} is sound.\qed
\end{proof}

\subsection*{Proposition~\ref{prop2}}

\section{Full Example}

  }

\end{document}